\documentclass[twocolumn]{aastex62}
\newcommand{\te}{$T_e$}    
\newcommand{\dne}{$n_e$}    
\newcommand{\ha}{H$\alpha$}    
\newcommand{\hb}{H$\beta$}    
\newcommand{\sii}{[S~{\sc ii}]}
\newcommand{\oiii}{[O~{\sc iii}]}
\newcommand{\nii}{[N~{\sc ii}]}

\newcommand{\hiireg}{{H{\sc ii}}}

\newcommand{\oii}{[O~{\sc ii}]}

\newcommand{\av}{$\rm A_V$}
\newcommand{\cmcube}{$\rm cm^{-3}$}
\newcommand{\ra}[1]{\renewcommand{\arraystretch}{#1}}

\graphicspath{{./}{figures/}}
\usepackage{amsmath}
\usepackage{booktabs}
\usepackage{CJK}
\usepackage{lineno}
\usepackage{longtable}

\received{January 1, 2018}
\revised{January 7, 2018}
\accepted{\today}
\submitjournal{ApJ}

\shorttitle{Nearby Simple HII Regions}
\shortauthors{Jin et al.}

\begin{document}
\begin{CJK*}{UTF8}{gbsn}

\title{Spatially-Resolved Temperature and Density Structures of Nearby \hiireg\ regions}

\correspondingauthor{Yifei Jin}
\email{Yifei.Jin@anu.edu.au}

\author[0000-0003-0401-3688]{Yifei Jin (金刈非)}
\affil{Research School for Astronomy \& Astrophysics, Australian National University, Canberra, Australia, 2611}
\affiliation{ARC Centre of Excellence for All Sky Astrophysics in 3 Dimensions (ASTRO 3D)}

\author{Ralph Sutherland}
\affiliation{Research School for Astronomy \& Astrophysics, Australian National University, Canberra, Australia, 2611}

\author{Lisa J. Kewley}
\affiliation{Research School for Astronomy \& Astrophysics, Australian National University, Canberra, Australia, 2611}
\affiliation{Institute for Theory and Computation, Harvard-Smithsonian Center for Astrophysics, Cambridge, MA 02138, USA}
\affiliation{ARC Centre of Excellence for All Sky Astrophysics in 3 Dimensions (ASTRO 3D)}

\author{David C. Nicholls}
\affiliation{Research School for Astronomy \& Astrophysics, Australian National University, Canberra, Australia, 2611}
\affiliation{ARC Centre of Excellence for All Sky Astrophysics in 3 Dimensions (ASTRO 3D)}

\begin{abstract}

Photoionization models frequently assume constant temperature or density within HII regions.  We investigate this assumption by measuring the detailed temperature and density structures of four \hiireg\ regions in the Large Magellanic Cloud and the Small Magellanic Cloud, using integral-field spectroscopic data from the Wide-Field Spectrograph on the ANU 2.3m telescope. 
We analyse the distribution of emission-lines of low-ionization species, intermediate-ionization species and high-ionization species. 
We present the complex electron temperature and density structures within \hiireg\ regions.
All four nebulae present a negative gradient in the electron density profile.
Both positive and negative temperature gradients are observed in the nebulae. 
We create a series of nebula models with a constant ISM pressure and varying temperature and density distributions.
Comparison of the line ratios between our \hiireg\ regions and models suggests that none of the simple nebula models can reproduce the observed temperature and density structures.
Comparison between the models and the data suggests that the ISM pressure of nebulae in LMC and SMC is between log($P/k$)=6-7.5.
Complex internal structures of the nebulae highlight the importance of future Monte-Carlo photoionization codes for accurate nebula modeling, which include a comprehensive consideration of arbitrary geometries of \hiireg\ regions.

\end{abstract}


\section{Introduction}\label{sec:intro}

To fully understand galaxies, we need to accurately model the \hiireg\ regions within galaxies.
The electron temperature and density are two of the fundamental quantities that determine the nebular emission-line spectrum within \hiireg\ regions.
Accurate predictions of the interstellar medium (ISM) physical quantities rely on the assumed or modeled distributions of electron temperature, \te , and density, \dne . 

The ISM electron temperature and density distributions across \hiireg\ regions are complex.
The density distributions in most \hiireg\ regions show filamentary or shell-like structures \citep{Kennicutt-1984}.
The density measurements within these structures have diverse radial gradients \citep{Binette-2002,Herrera-Camus-2016,Rubin-2016}, although relatively uniform density distributions are seen in some cases \citep{Garnett-2001,Krabbe-2002,Garcia-Benito-2010}. 
Surveys of both galactic and extragalactic \hiireg\ regions indicate that approximately half the \hiireg\ regions have internal variations in electron density \citep{Copetti-2000,Malmann-2002}.

The presence of temperature fluctuations in \hiireg\ regions is under significant debate due to the notorious difficulty in measuring the electron temperature.
Some studies show evident temperature fluctuations \citep{Peimbert-2004,Hagele-2006,Kewley-2019b},
while others find no temperature fluctuations \citep{Stasinska-2013,Liu-2006}. 

The underlying temperature and density structures in \hiireg\ regions affect the robustness of metallicity calibrations.
The most direct calibration method is to convert metallicities from electron temperatures in the $\rm O^{2+}$ zone after correcting for the unseen ionization stages (the so-called ionization correction factor, ICF).
The ICF is determined from the photoionization models under the assumption of constant electron temperature and density across the \hiireg\ region.
\cite{Hagele-2008} find electron temperature variations raise the uncertainty of metallicities by up to 0.2 dex through the ICF method.

The temperature variation is one of the potential causes of the discrepancies among metallicities measured from different emission-lines \citep{Kewley-2019b}.
Commonly used metallicity-sensitive emission-line ratios are (\oii $\lambda 3727$ + \oiii $\lambda\lambda 4959,5007$)/\hb\ (known as $R_{23}$), \oiii $\lambda5007$/\hb/\nii $\lambda6584$/\ha (known as O3N2) and \nii $\lambda6584$/\ha.
Recombination lines give larger metallicities than auroral lines. 
The offset between recombination line and auroral metallicities has been suggested as arising from temperature fluctuations in \hiireg\ regions \citep{Peimbert-1967}, which is known as the abundance discrepancy factor (ADF).

The ISM pressure incorporates the complex structures of electron temperature and density.
Constant pressure models are applicable to the majority of \hiireg\ regions where the sound-crossing time is shorter than the heating and cooling timescale.
Pressure photoionization models with complex temperature and density structures have been successfully applied to galaxies at $z\sim2$, suggesting the existence of the high-pressure \hiireg\ regions in high-$z$ galaxies \citep{Lehnert-2009}.

Current photoionization models significantly underestimate the complexity of the internal structures of \hiireg\ regions.
Isothermal or constant density ISM conditions are often assumed in photoionization models.
Isothermal models with a single temperature across the nebula have been used in calibrating the standard density-sensitive line ratios \citep{Osterbrock-1989,Rubin-1994}, and determining the excitation power source in galaxies \citep{Filippenko-1985,Groves-2004}.
The constant density distribution assumption is adopted in measuring the electron temperature and density in \hiireg\ regions \citep{Osterbrock-1989}.
The constant density assumption is also used in deriving the metallicities measured from auroral lines.
In modern constant pressure models, the majority of modeled \hiireg\ regions produce complex internal gradients in electron temperature and density, while only few specific cases with log(O/H)+12 $\sim 8.23$ can be approximated with the isothermal or constant density conditions \citep{Kewley-2019}.

Nearby \hiireg\ regions are laboratories for understanding the internal structure of the ISM electron temperature and density.
With the advantage of proximity, the Magellanic Clouds provide a rich reservoir of \hiireg\ regions for better understanding the ISM temperature and density fluctuations \citep{Kennicutt-1984,Kennicutt-1986,Chu-1986,Vermeij-2002,Pena-Guerrero-2012}.
Integral-field unit (IFU) observations offers highly spatially-resolved data of \hiireg\ regions in the Magellanic Clouds for the detailed study of electron temperature and density structures \citep{Dopita-2019,Kumari-2017}. 
Future IFU data sets of nearby \hiireg\ regions will be released in the on-going TYPHOON program (Seibert et al. in prep.) and the up-coming SDSS-V/Local Volume Mapper (LVM) \citep{Kollmeier-2017}. 

In this paper, we present deep IFU observations of four \hiireg\ regions in the Large Magellanic Cloud (LMC) and the Small Magellanic Cloud (SMC).
The data were obtained using the Wide-Field Spectrograph (WiFeS) with grating resolution of $R\sim7000$.
We derive the electron temperature and density with the deep IFU data, and further analyze the thermal and density structures within each \hiireg\ region using the highly spatially-resolved datacube.
This is the first study of nebular internal structures with a sample of LMC/SMC nebulae using the IFU data with a resolution of $\rm 0.2-0.3~pc$.
This paper is structured as below.
We first describe the sample of \hiireg\ regions in Section~\ref{sec:sample} and describe the data and the data reduction in Section~\ref{sec:data}.
We describe the calibrations of electron temperature, density and pressure in Section~\ref{sec:quant}.
We present the electron temperature and density structures of \hiireg\ regions in Section~\ref{sec:analy}.
In Section~\ref{sec:model}, we compare our \hiireg\ regions with the pressure photoionization models.
The discussion is given in Section~\ref{sec:dis}.

\section{The sample}\label{sec:sample}

\begin{figure*}
  \centering
  \includegraphics[width=7in]{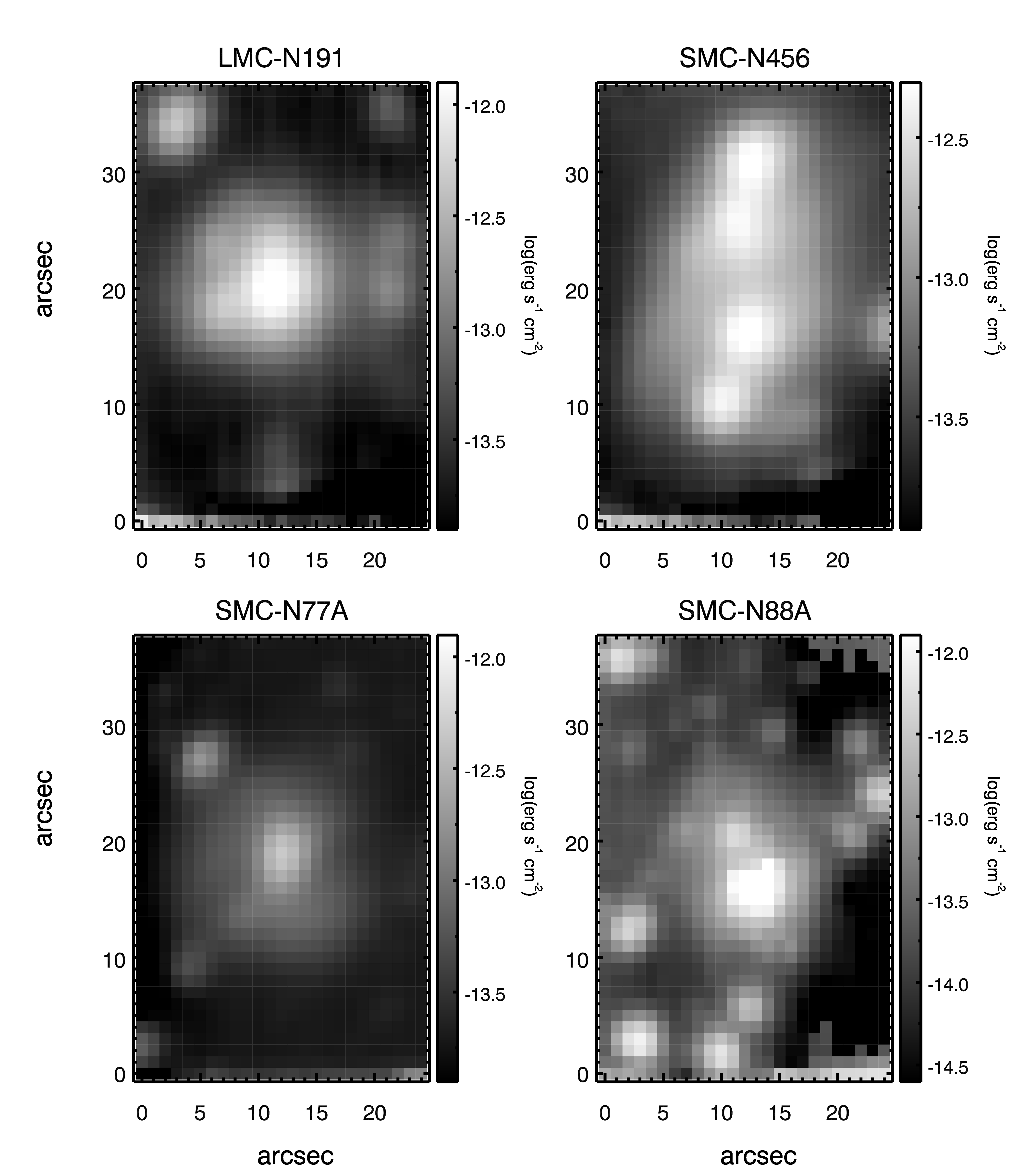}
  \caption{Collapsed IFU images of four \hiireg\ regions in LMC and SMC. The field of view is $25^{\prime\prime}\times38^{\prime\prime}$. Grey scale color indicates the line-of-sight integrated flux. }\label{fig:map}
\end{figure*}

The four \hiireg\ regions are selected from a series of spatially-resolved observations of the nebulae in the Magellanic Clouds \citep{Dopita-2016,Dopita-2018A}.
The four \hiireg\ regions are bright and young compact \hiireg\ regions ionized by a single O-star or simple star clusters.
Three \hiireg\ regions, N456, N77A and N88A are from SMC, and one, N191A, is from LMC.
The simple ionizing sources make these compact \hiireg\ regions ideal targets for testing current photoionization models.
The compact size ($\sim10^{\prime\prime}$ in diameter) and proximity allows these compact \hiireg\ regions to fit within the WiFeS field of the view (FoV).
Some fundamental physical quantities derived from the global spectra of these \hiireg\ regions are listed in Table~\ref{tab:samp}.

LMC~N191 is a complex \hiireg\ region which presents two components in \ha , \oiii\ and \hb\ narrow band images \citep{Henize-1956}.
N191A is more compact and brighter between the two components, which has a diameter of $5^{\prime\prime}.2$ (1.2 pc at LMC distance) \citep{Selier-2012}.
The broad $V$-band image reveals that the ionized gas in N191A is excited by an O-star located at the center of the nebula \citep{Selier-2012}.

SMC~N77A is a spherical \hiireg\ region having a radius of $10^{\prime\prime}$ \citep{Selier-2012}. 
A dust lane crosses SMC~N77A along the east-west direction \citep{Selier-2012,Toribio-2017}.
The major ionizing source of N77A is a B-star residing north of the dust lane. 
The average extinction of N77A is \av$\sim0.25$~mag \citep{Selier-2012}. 

SMC~N88A is the brightest \hiireg\ region in the SMC \citep{Testor-2003}.
N88A is a $\sim1$-pc sized ionized gas ``blob'' excited by an O3-4 star \citep{Testor-2003}.
The heavy dust core in N88A produces its high integral extinction of \av$ = 1.38$~mag \citep{Garnett-1995,Testor-2003}.

SMC~N456 is the second brightest \hiireg\ region in the SMC \citep{Peimbert-1976}, with an inhomogeneous thermal structure.
Long-slit spectrographic study indicates the mean-square temperature variation, $t^{2}$ (first introduced by \cite{Peimbert-1967}), in SMC~N456 is $0.067\pm0.013$ \citep{Pena-Guerrero-2012a}.

\begin{deluxetable}{lccccc}
\tablewidth{8truecm}
\tablecaption {The fundamental physics quantities of the \hiireg\ region sample \tablenotemark{a}}
\tablehead {
\colhead {\hiireg\ Region} & \colhead {Size} & \colhead {$\rm A_V$} & \colhead {$T_e$} & \colhead {$n_e$} \\
\colhead {} & \colhead {pc} & \colhead {mag} & \colhead {K} & \colhead {cm$^{-3}$} \\ 
}
\startdata  
\ LMC N191A & 1.2  &     $\sim2.70$  & 10,800 & 600  \\
\ SMC N77A  &  2.9  &     $\sim0.25$  & 14,240 & 60  \\
\ SMC N88A  &  1.2  &     $\sim1.38$  & $\sim14,100$  & 2870   \\
\ SMC N456&   --  &     --  & $\sim12,000$ & $\sim100$  \\
\enddata  
\tablenotetext{a}{The data are collected from the literature \cite{Selier-2012} for LMC N191A and SMC N77A, \cite{Testor-2003} for SMC N88A and \cite{Pena-Guerrero-2012a} for SMC N456.}
\end{deluxetable}\label{tab:samp}

\section{Observation and data reduction}\label{sec:data}

The spectroscopic data were observed using WiFeS.
WiFeS \citep{Dopita-2007,Dopita-2010} is an image-slicing integral-field spectrograph mounted at the Australian National University 2.3 m telescope at the Siding Spring Observatory.
WiFeS has a $25^{\prime\prime} \times\ 38^{\prime\prime} $ field of the view with spaxels each $1 \times\ 1^{\prime\prime}$ in angular size.
The double-beam spectrograph feeds the light into the blue and red cameras simultaneously to cover the wavelength ranging from 3300 to 9000 \AA .
We used the high-resolution gratings of $R=7000$ by splitting the spectra into $U$, $B$, $R$ and $I$ bands. 

The observations were taken between November and December 2015 with seeing FWHM 1.0-2.0 arcsec.
Each \hiireg\ region was observed using 10, 100 and 1000 (or 800) second exposures. 
The longer exposures were aimed at securing flux measurements of faint emission lines down to $\sim10^{-5}$ of the \hb\ flux. 
The shorter exposures were necessary to minimize the saturation on the bright hydrogen and oxygen lines.

The data were reduced using the standard WiFeS pipeline, PyWiFeS \citep{Childress-2014} . 
Telluric lines were removed by subtracting observed clear sky regions adjacent to the \hiireg\ regions, from each observation. 
The separate camera spectra were then combined using the overlapping spectral regions between adjacent cameras.
The final data has a spectral resolution of 43~$\rm km s^{-1}$ and a spatial resolution of 0.3~pc for SMC N77A, SMC N88A and SMC N456, and a spatial resolution of 0.24~pc for LMC N191A.

\section{Derived quantities}\label{sec:quant}

\subsection{Emission-line maps}

\begin{figure*}
  \centering
  \includegraphics[width=7in]{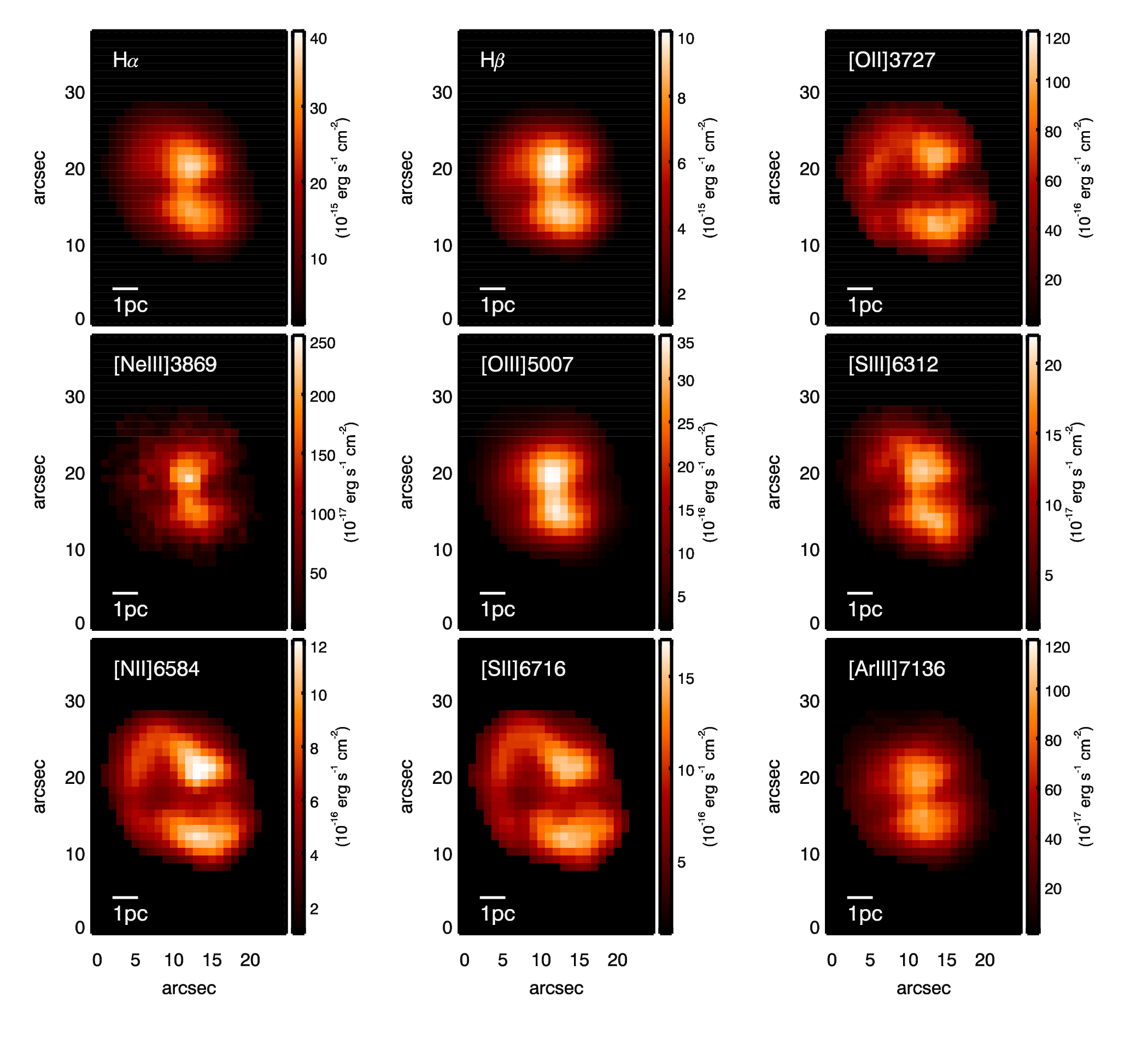}
  \caption{Emission-line maps of SMC~N77A. We show the maps of Balmer lines \ha\ and \hb , the emission-lines of high-ionization species \oiii$\lambda5007$ and [Ne~{\sc iii}]$\lambda3869$, the emission-lines of intermediate-ionization species [Ar~{\sc iii}]$\lambda7136$, [S~{\sc iii}]$\lambda6312$, and the emission-lines of low-ionization species [O~{\sc ii}]$\lambda3727$, [N~{\sc ii}]$\lambda6584$ and [S~{\sc ii}]$\lambda6716$.}\label{fig:elmap_smcn77a}
\end{figure*}

\begin{figure*}
  \centering
  \includegraphics[width=7in]{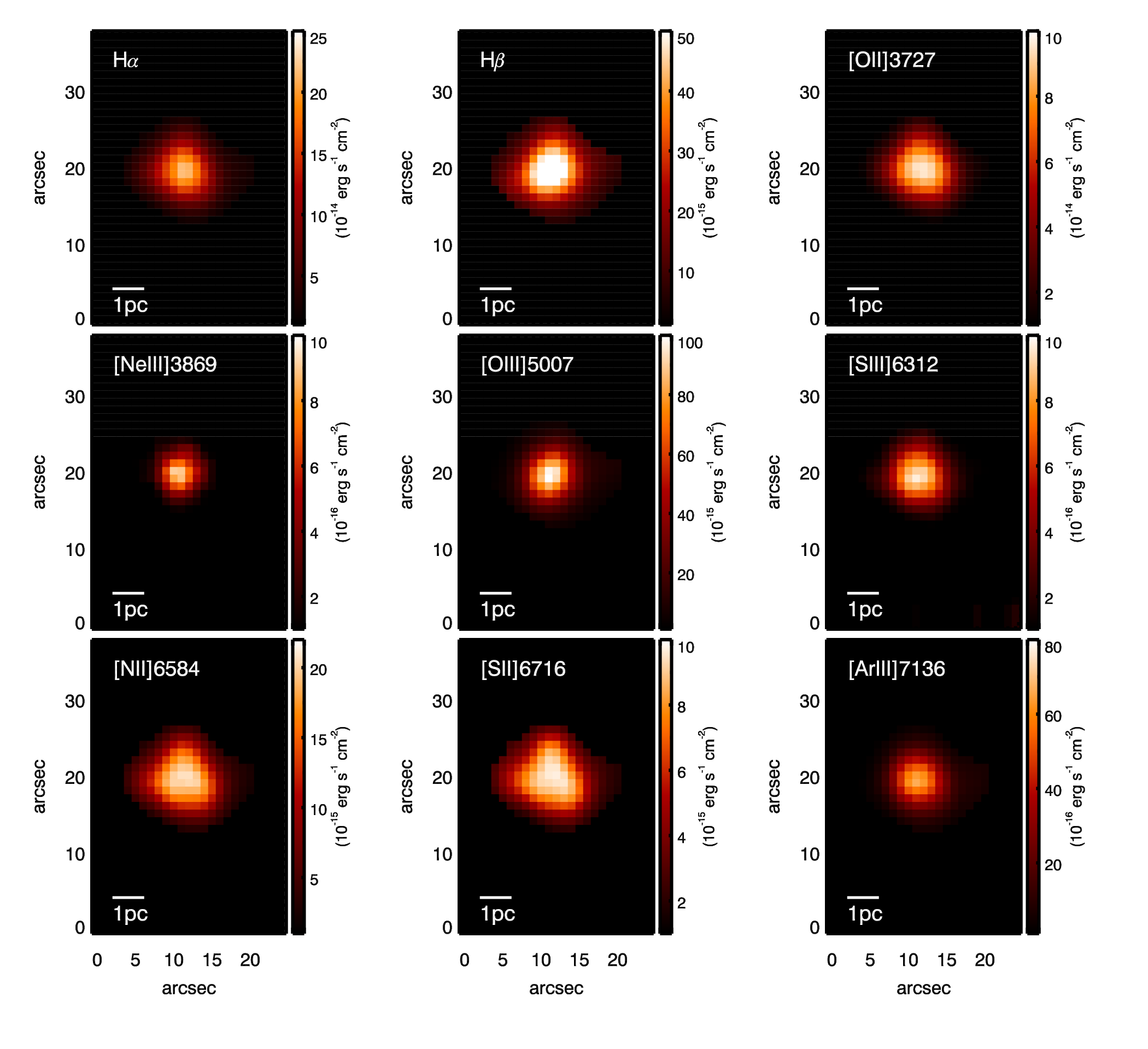}
  \caption{Same as Figure~\ref{fig:elmap_smcn77a} but for LMC~N191.}\label{fig:elmap_lmcn191}
\end{figure*}

\begin{figure*}
  \centering
  \includegraphics[width=7in]{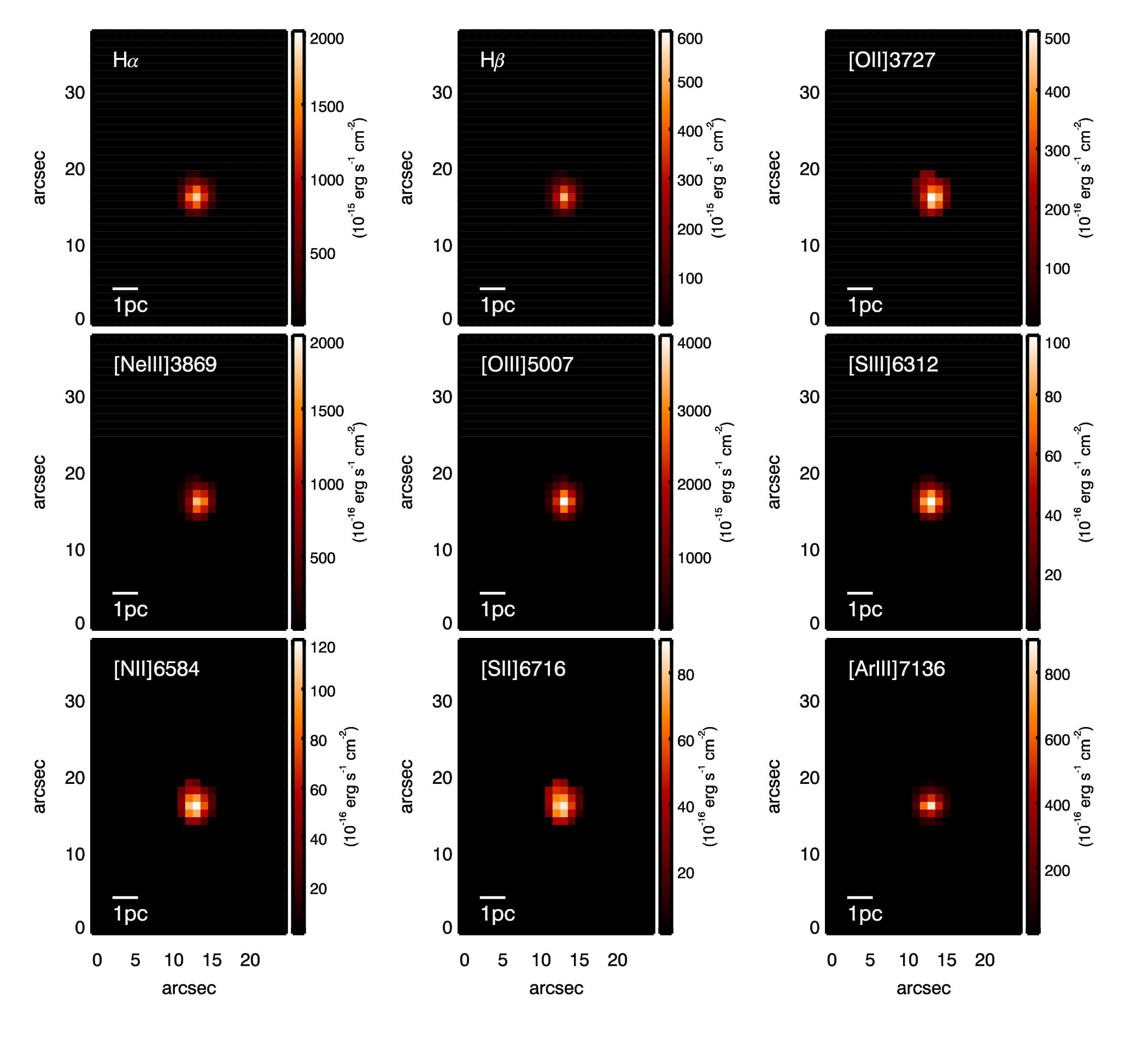}
  \caption{Same as Figure~\ref{fig:elmap_smcn77a} but for SMC~N88A.}\label{fig:elmap_smcn88a}
\end{figure*}

\begin{figure*}
  \centering
  \includegraphics[width=7in]{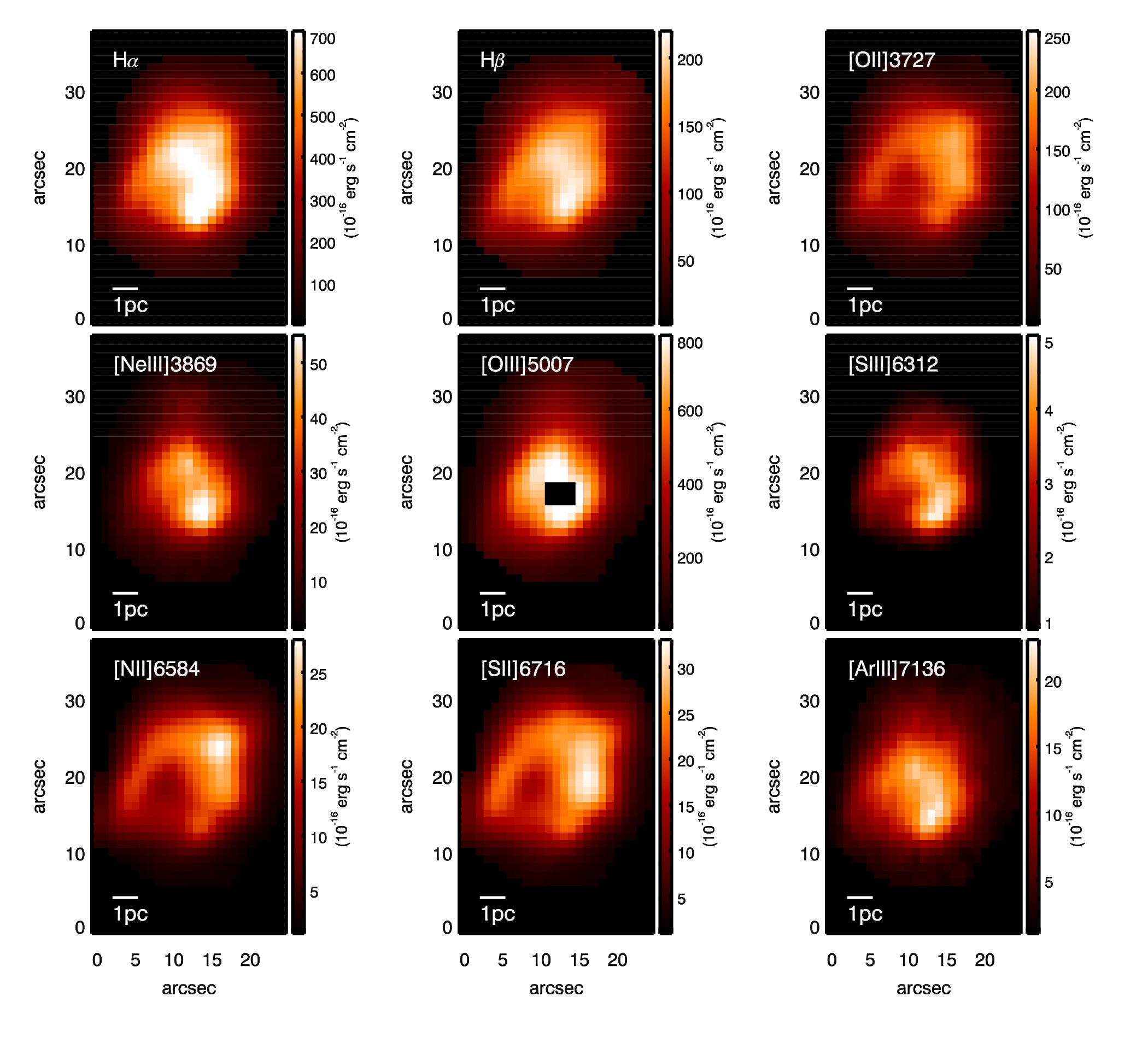}
  \caption{Same as Figure~\ref{fig:elmap_smcn77a} but for SMC~N456. The central pixel in the \oiii$\lambda$5007 map is a bad pixel because the \oiii$\lambda$5007 is saturated at the center of SMC~N456.}\label{fig:elmap_smcn456}
\end{figure*}


The emission-line fluxes are obtained from each pixel by integrating the spectrum subtracted from the underlying stellar continuum and nebular continuum.
Continuum components are estimated by using the average flux over a 10\AA\ wavelength interval on either the left side or the right side to each emission-line. 
In Figure~\ref{fig:elmap_smcn77a} to Figure~\ref{fig:elmap_smcn456}, we show maps of the emission-lines representing species in the high-ionization zone, the intermediate-ionization zone and the low-ionization zone of the nebula.
We use the Balmer lines to trace the main body of \hiireg\ regions.
The \oiii$\lambda5007$ and {[Ne \sc iii]}$\lambda3869$ lines are used to trace the high-ionization zone.
The {[Ar \sc iii]}$\lambda7136$ and {[S \sc iii]}$\lambda6312$ lines are the tracers of the intermediate-ionization zone.
The \oii$\lambda3727$, \nii$\lambda6584$ and {[S \sc ii]}$\lambda6716$ emission-lines are used to trace the low-ionization zone of the nebula.
The boundary of the \hiireg\ region is identified at the pixels with $\rm Flux(H\alpha)=$10\%$\rm Flux(H\alpha)_{peak}$, where $\rm Flux(H\alpha)_{peak}$ is the maximal \ha\ flux.

In Figure~\ref{fig:elmap_smcn77a} to Figure~\ref{fig:elmap_smcn456}, a clear spatial segregation exists among different ionization zones.
The high-ionization zone that is traced by \oiii$\lambda5007$ and {[Ne \sc iii]}$\lambda3869$ is compact around the nebular center because of the hard ionization field from the ionizing star in proximity to the star.
In contrast, the low-ionization zone traced by \oii$\lambda3727$, \nii$\lambda6584$ and {[S \sc ii]}$\lambda6716$ is located at the boundary of the \hiireg\ regions.
The {[Ar \sc iii]}$\lambda7136$ and {[S \sc iii]}$\lambda6312$ lines are located between the high-ionization and low-ionization zones.

Our \hiireg\ region sample presents diverse nebular morphologies.
The emission-lines in LMC~N191 have an extended morphology, which is consistent with the early \ha\ image observed by \cite{Henize-1956}. 
SMC~N88A appears as a bright compact point source for the emission-lines of high-ionization and intermediate-ionization species, which is consistent with the near-infrared imaging of SMC~N88A \citep{Testor-2010}.
For the emission-lines of low-ionization species, an extended region exists north-east of the bright nebular center.   
SMC~N456 presents a shell-like morphology for the emission-lines of low-ionization species. 
However, the emission-lines of high-ionization species present a spherical morphology.
The extended morphology of SMC~N456 is also confirmed in \cite{Arellano-2020}.
SMC~N77A appears to have a two-lobe morphology for emission-lines of high-ionization and intermediate-ionization species, which is split by a dust lane crossing from east to west \citep{Selier-2012}.
For the low-ionization species, the emission-lines are located on the nebular edge, leading to a shell-like morphology.

\subsection{Extinction Correction}

\begin{figure*}
  \centering
  \includegraphics[width=7in]{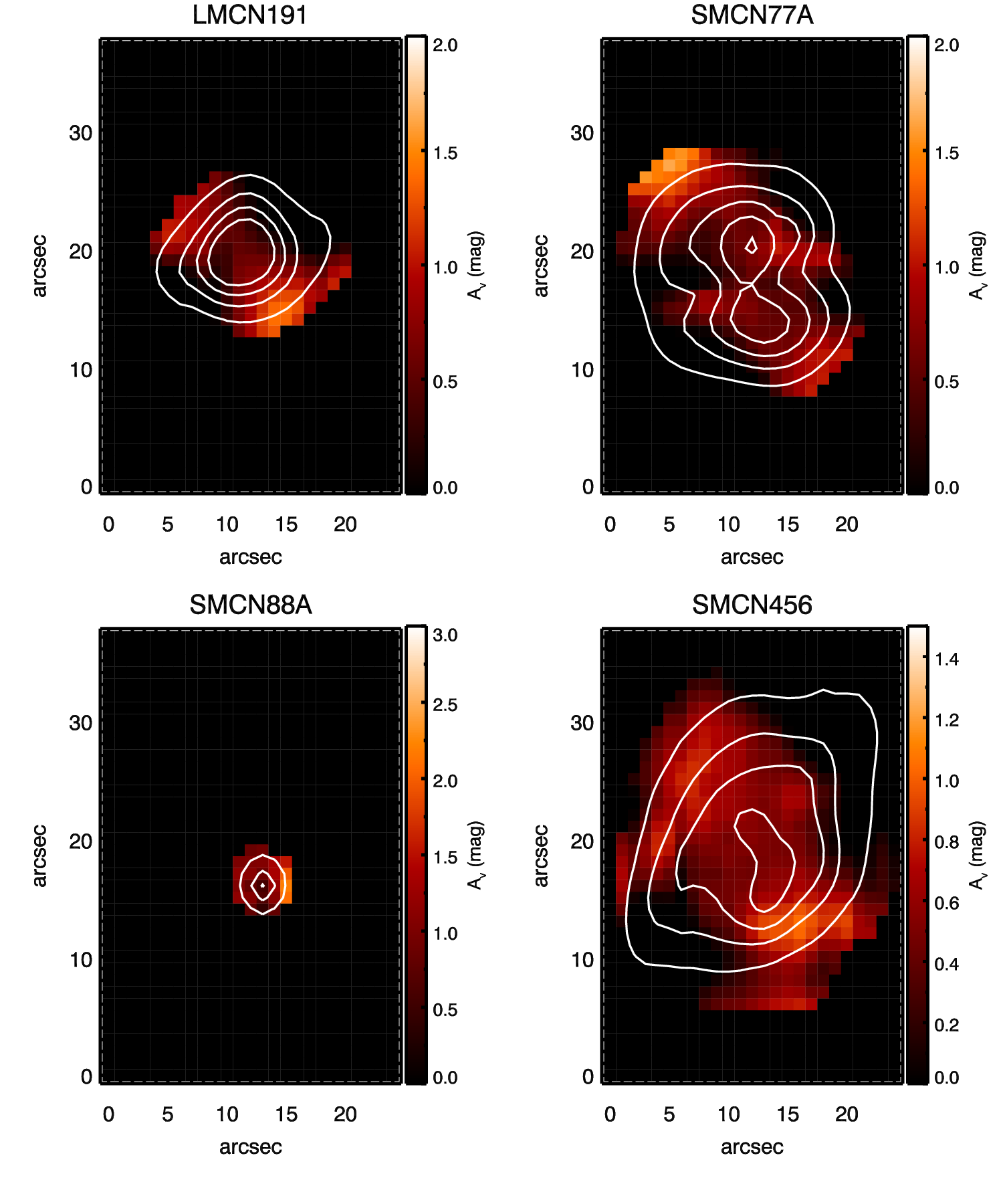}
  \caption{Maps of attenuation \av. The color code indicates the value of \av . The white contours present the distribution of \hb\ emission-line fluxes.}\label{fig:avmap}
\end{figure*}

The extinction correction is carried out through the \ha/\hb\ Balmer decrement, using the \cite{Fitzpatrick-1999} extinction law.
We select a Rv$ = 3.16$ for nebula in LMC and Rv$ =  2.93$ for nebula in SMC \citep{Pei-1992}. 
The Fitzpatrick extinction curve is a generally-used parameterized curve. 
Compared to the average extinction curves of the LMC and the SMC \citep{Gordon-2003}, we find the difference of the extinction corrected fluxes is 0.7\%-8\% by applying different extinction curves on our sample.

Figure~\ref{fig:avmap} presents maps of the derived \av\ with the distribution of \hb\ fluxes.
We notice that the enhancement of \av$\sim0.7$~$mag$ in the SMC~N77A is at the position of the central dust lane.
In SMC~N88A, the maximum \av\ is $\sim$2.5~$mag$, located at the peak of \hb\ fluxes.

\subsection{Determining electron density}

We measure the electron density, $n_e$, for each resolution element by using the calibrations in \cite{Osterbrock-2006} (hereafter OF06) based on single-atom fixed temperature, fixed density models, and the diagnostics given by \cite{Kewley-2019} (hereafter K19) based on their theoretical density models.
Compared with single-atom models, the density models in \cite{Kewley-2019} are more representative to the physical conditions in nebulae in terms of the temperature structures changing with the ionization field and the metallicity.
The electron densities derived from the \cite{Osterbrock-2006} method and the \cite{Kewley-2019} density models are independent from the ISM pressure, which can be used to compare with the constant pressure models where the ISM pressure is a free parameter.

The electron densities are derived from the \sii$\lambda6716$/ \sii$\lambda6731$ ratio and the \oii$\lambda3726$/ \oii$\lambda3729$ ratio respectively.
We use the amplitude-to-noise (A/N) ratio to describe the quality of each emission-line.
The A/N ratio is the ratio of the peak of emission-line flux to the noise of emission-lines, tracing both the signal-to-noise ratio and the strength of emission-line fluxes.
The A/N ratio is sensitive to the emission-lines with low signal-to-noise ratios but evident excess of fluxes superimposed on the continuum.
We require A/N of the \oii\ doublets and the \sii\ doublets larger than 5 in each spaxel.

The OF06 models use the atomic data from \cite{Mendoza-1983}, which provides a compilation of transition probabilities, electron excitation coefficients and photoionization cross sections from sophisticated computation and measurements.
The K19 models use the CHIANTI v.8.0 atomic data \citep{DelZanna-2015} in conjunction with the energy level information of each ions from NIST database. 
The K19 models are suitable for calculating the electron density between 1 and $\rm10^8~cm^{-3}$.

\subsection{Determining electron temperature}

We measure the electron temperatures, $T_e$, from collisional lines using the method described in \cite{Osterbrock-2006} and the calibration in \cite{Nicholls-2020} (hereafter NKS20).
The \cite{Osterbrock-2006} method is a simple polynomial fit to the single-atom models within a temperature range from 5000~K to 25,000~K.
\cite{Nicholls-2020} re-calibrate the relationship between the \oiii\ emission-line ratios and $T_e$ by using a rational polynomial as shown in equation~\ref{eq:nicholls-fit} 
\begin{equation}
{\rm log}_{10}(T_e) = \frac{P_0 + P_1x + P_2x^2}{1 + Q_1x + Q_2x^2 + Q_3x^3},
\label{eq:nicholls-fit}
\end{equation}
where $x$ is $\rm log_{10}${(\oiii$\lambda4363$/\oiii$\lambda\lambda4959,5007$)}, and $P_0$, $P_1$, $P_2$, $Q_1$, $Q_2$, $Q_3$ are the fit coefficients.
We require A/N$>$5 for \oiii $\lambda\lambda$4959,5007 and \oiii $\lambda$4363

The OF06 models use the atomic data from \cite{Mendoza-1983}.
The NKS20 models use the CHIANTI v.8.0 atomic data \citep{DelZanna-2015} in conjunction with the energy level information of each ions from NIST database.

\subsection{Determining $P/k$}

The pressure within \hiireg\ regions is derived from the temperature, $T$, and the particle density, $n$, of nebula through the ideal gas assumption $P=nkT$.
We assume that $T\equiv T_e$ and approximate the total particle density $n$ from the electron density $n_e$ through $n\sim2 n_e (1+He/H)$ given hydrogen and helium are the majority species in \hiireg\ regions.
Finally, the nebular pressure is derived through $P/k = 2 n_e (1+He/H)T_{e}$.

\section{Analysis}\label{sec:analy}


\begin{figure*}
  \centering
  \includegraphics[width=7in]{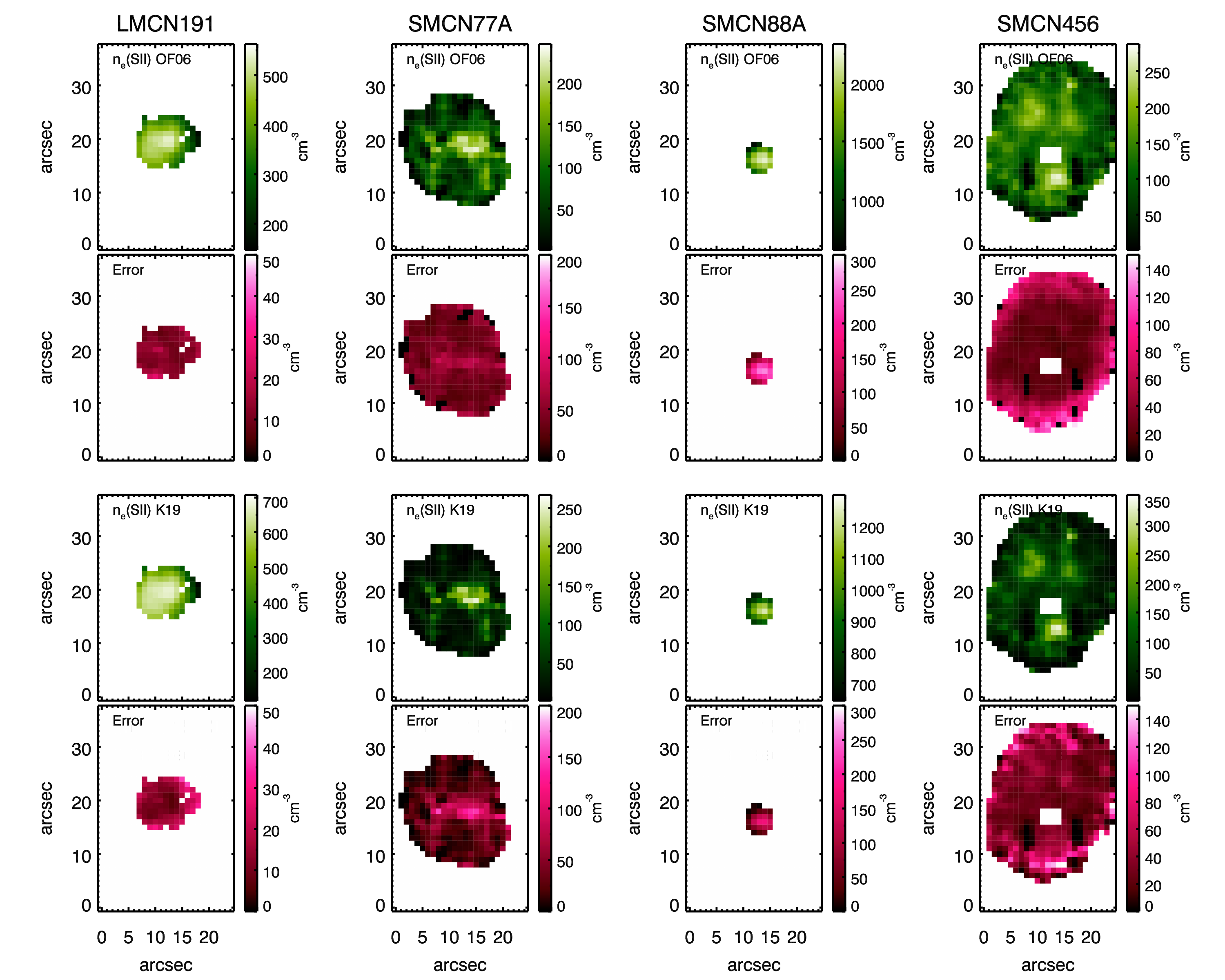}
  \caption{Maps of electron density $n_e$. 
  The color code indicates the value of $n_e$.
  {\bf First row:} Maps of the electron density derived from the \sii\ ratio based on the method in \cite{Osterbrock-2006}.
  {\bf Second row:} Maps of errors of the electron density derived from the \sii\ ratio based on the method in \cite{Osterbrock-2006}.
  {\bf Third row:} Maps of the electron density derived from the \sii\ ratio based on the method in \cite{Kewley-2019}.
  {\bf Fourth row:} Maps of errors of the electron density derived from the \sii\ ratio based on the method in \cite{Kewley-2019}.}\label{fig:nemap}
\end{figure*}

\begin{figure*}
  \centering
  \includegraphics[width=7in]{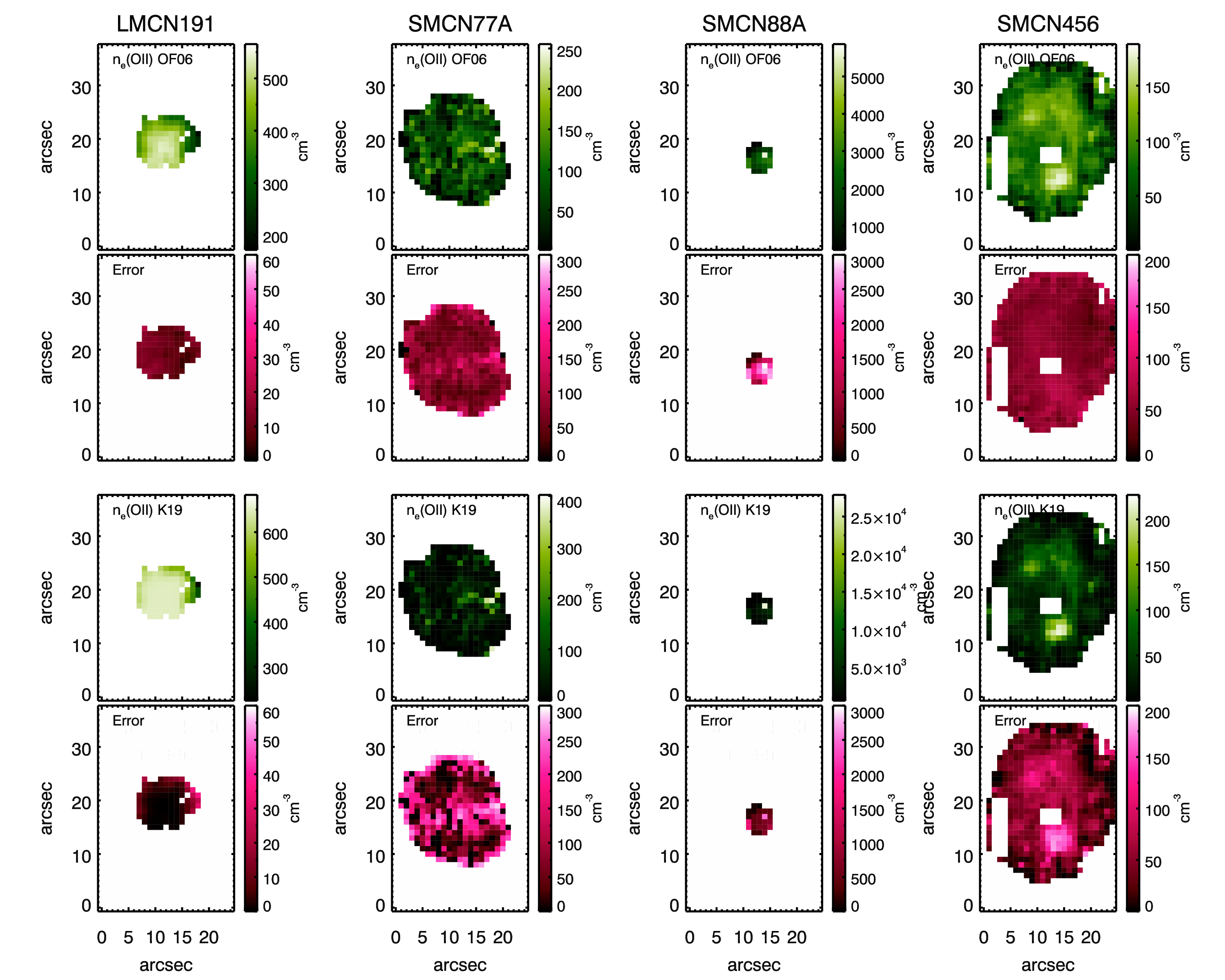}
  \caption{Maps of electron density $n_e$. 
  The color code indicates the value of $n_e$.
  {\bf First row:} Maps of the electron density derived from the \oii\ ratio based on the method in \cite{Osterbrock-2006}.
  {\bf Second row:} Maps of errors of the electron density derived from the \oii\ ratio based on the method in \cite{Osterbrock-2006}.
  {\bf Third row:} Maps of the electron density derived from the \oii\ ratio based on the method in \cite{Kewley-2019}.
  {\bf Fourth row:} Maps of errors of the electron density derived from the \oii\ ratio based on the method in \cite{Kewley-2019}.}\label{fig:nemap2}
\end{figure*}

\begin{figure*}
  \centering
  \includegraphics[width=7in]{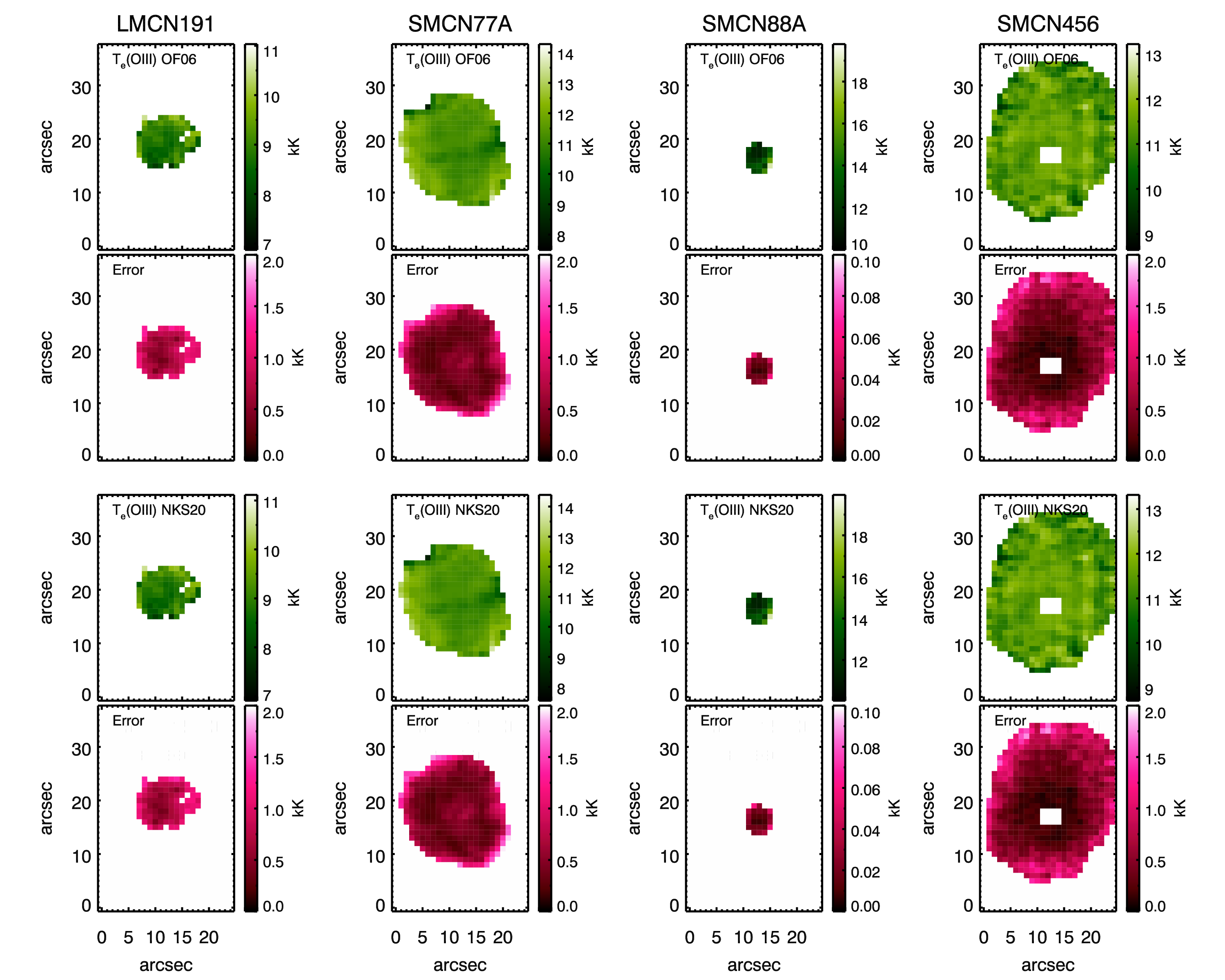}
  \caption{Maps of electron temperature \te. 
  The color code indicates the value of \te . 
  {\bf First row:} Maps of electron temperature derived from the \oiii\ ratio based on the method in \cite{Osterbrock-2006}.
  {\bf Second row:} Maps of the error of electron temperature measured from the \oiii\ ratio based on the method in \cite{Osterbrock-2006}.
  {\bf Third row:} Maps of electron temperature derived from the \oiii\ ratio based on the method in \cite{Nicholls-2020}.
  {\bf Fourth row:} Maps of the error of electron temperature measured from the \oiii\ ratio based on the method in \cite{Nicholls-2020}.}\label{fig:temap}
\end{figure*}


\subsection{Electron density structures}

In Figure~\ref{fig:nemap} and Figure~\ref{fig:nemap2}, we present the maps of electron density derived from the \oii\ ratio, $n_e(\rm OII)$, and the \sii\ ratio, $n_e(\rm SII)$. 
In LMC~N191, SMC~N77A and SMC~N88A, the 2D density distribution has a dense core at the center of the nebula and the density decreases along the radius of nebula.
In SMC~N456, there are three dense clumps located at three corners of the nebula.

We measure the gradients of the electron density profiles using the following function,
\begin{equation}
\frac{n_e}{cm^{-3}} = \alpha + \beta \left( \frac{r}{pc} \right),
\label{eq:fit_func}
\end{equation}
where $\beta$ is the gradient of the profile.
The center of nebula is identified as the position of the peak flux of \ha .
The best-fit parameters are listed in Table~\ref{tab:fits}.

The derived density gradient changes by using different methods of the measurement.
LMC~N191 and SMC~N88A have evident negative density gradients with a slope of -75~$\rm cm^{-3}~pc^{-1}$$<\beta< $-45$\rm cm^{-3}~pc^{-1}$ for LMC~N191 and -850~$\rm cm^{-3}~pc^{-1}$$<\beta< $-175 $\rm cm^{-3}~pc^{-1}$ for SMC~N88A.
In LMC~N191, the density gradient derived from the K19 \oii-ratio method is smaller than the gradient derived from other methods. 
In contrast, the SMC~N77A and SMC~N456 have the shallow and even flat density gradients.
In SMC~N77A, the density gradient derived from the \sii-ratio is marginally negative by -8$\pm$5 $\rm cm^{-3}~pc^{-1}$$<\beta< $-6$\pm$5 $\rm cm^{-3}~pc^{-1}$.
The gradient derived from the \oii-ratio is flat with a $\beta\sim$-4$\pm$5.5 $\rm cm^{-3}~pc^{-1}$.
In SMC~N456, the density gradient is flat within the error range by $\beta\sim$-3$\pm$3 $\rm cm^{-3}~pc^{-1}$.

\begin{table*}[!t]
\centering
\caption{The average and standard deviation of the temperature by pixels of each \hiireg\ region\\ }
\label{tab:avg_temp}
\begin{tabular}{*{5}c}
\hline
\hline
(K)       &  LMCN191& SMCN77A & SMCN88A & SMCN456 \\
\hline
 $< T_{e} >_{OF06}$            & \multicolumn{1}{r}{8864}    & \multicolumn{1}{r}{11405}  & \multicolumn{1}{r}{12932}   & \multicolumn{1}{r}{11321} \\
 $\sigma(T_{e})_{OF06}$        & \multicolumn{1}{r}{813}     & \multicolumn{1}{r}{630}    & \multicolumn{1}{r}{1763}    & \multicolumn{1}{r}{423}     \\
 $< T_{e} >_{NKS20}>$          & \multicolumn{1}{r}{8923}    & \multicolumn{1}{r}{11490}  & \multicolumn{1}{r}{13030}   & \multicolumn{1}{r}{11405}   \\
 $\sigma(T_{e})_{NKS20}$       & \multicolumn{1}{r}{823}     & \multicolumn{1}{r}{636}    & \multicolumn{1}{r}{1779}    & \multicolumn{1}{r}{427}      \\
 \hline
\end{tabular}

\end{table*}

\subsection{Temperature structures}

In Figure~\ref{fig:temap}, we present the maps of electron temperatures derived from \oiii-ratio through the methods given by \cite{Osterbrock-2006} and \citep{Nicholls-2020}.
The temperature is evenly distributed across the entire nebula for the four \hiireg\ regions.

We calculate the average and the standard deviation of $T_e$ of each \hiireg\ region.
As shown in Table~\ref{tab:avg_temp}, the average temperature of the \hiireg\ region is around 8\,000~K to 13\,000~K, which is consistent with the previous measurement from the integrated spectra \citep{Testor-2003,Pena-Guerrero-2012a,Selier-2012}.
In each nebula, the measured temperature is consistent between the OF06 method and the NKS20 method.

We derive the temperature gradient of each \hiireg\ region by fitting temperature radial profiles with an analytical function as below,
\begin{equation}
\frac{T_e}{K} = \alpha + \beta \left( \frac{r}{pc} \right),
\label{eq:fit_func}
\end{equation}
where $r$ is the radius of nebula.
The center of nebula is identified as the position of the peak flux of \ha .
The best-fit parameters are listed in table~\ref{tab:fits}.

The temperature gradient is flat in the four \hiireg\ regions.
The gradient is negligible to the average temperature of each nebula by $\beta$~$\approx$~5\%$<T_e>$, where $<T_e>$ is the measured average temperature.
LMC~N191, SMC~N77A and SMC~N88A show a positive trend in the temperature radial profile with the gradient ranging from 37$\pm$41 K~$\rm pc^{-1}$ to 180$\pm$320 K~$\rm pc^{-1}$.
SMC~N456 shows a negative trend in the temperature radial profile with a gradient of -27$\pm$30 K~$\rm pc^{-1}$.
The gradient values are smaller than the fitting errors in the four nebulae.

\begin{table*}
\ra{1.3}
\caption{Parameters of the best-fits to the radial profiles of electron temperature and electron density$^a$.}
\begin{tabular}{@{}rrrcrrcrrcrr@{}}\toprule
\centering
&\multicolumn{2}{c}{LMC N191}& \phantom{abc} & \multicolumn{2}{c}{SMC N77A} & \phantom{abc} & \multicolumn{2}{c}{SMC N88A} & \phantom{abc} & \multicolumn{2}{c}{SMC N456} \\
\cmidrule{2-3} \cmidrule{5-6} \cmidrule{8-9} \cmidrule{11-12} 
& {$\alpha$} & $\beta$ && $\alpha$ & $\beta$ && $\alpha$ & $\beta$ && $\alpha$ & $\beta$\\ 
\midrule
\midrule
$T_e(OIII)$\\
(OF06)& 8570$\pm$256 & 73$\pm$87 &&   11117$\pm$156  &  37$\pm$41  &&  12299$\pm$7  & 179$\pm$317        &&  11610$\pm$252 & -27$\pm$30\\
\cmidrule{1-12}
$T_e(OIII)$\\
(NKS20)& 8626$\pm$259 & 75$\pm$88 &&  11200$\pm$157  &  37$\pm$41  &&  12391$\pm$8  & 181$\pm$320        &&  11697$\pm$253 & -27$\pm$31\\
\midrule
\midrule
$n_e(SII)$\\
(OF06)& 582$\pm$21    & -54$\pm$5 &&  143$\pm$49     &  -8$\pm$5   &&  2375$\pm$179 & -396$\pm$117       &&  156$\pm$45 & -3$\pm$3\\
\cmidrule{1-12}
$n_e(SII)$\\
(K19)& 748$\pm$17    & -72$\pm$6  &&   88$\pm$51     &  -6$\pm$5   &&  1276$\pm$96  & -172$\pm$40         &&  112$\pm$52 & -3$\pm$3\\
\midrule
\midrule
$n_e(OII)$\\
(OF06)& 579$\pm$13   & -46$\pm$5  &&  107$\pm$37     & -4$\pm$5    &&  3033$\pm$717 & -722$\pm$353       &&  115$\pm$30 & -3$\pm$3\\
\cmidrule{1-12}
$n_e(OII)$\\
(K19)& 655$\pm$2    & -5$\pm$2   &&   73$\pm$54        &  -4$\pm$6  && 3244$\pm$483 & -831$\pm$189        &&  61$\pm$38 & -2$\pm$4\\
\midrule
      
\bottomrule
\end{tabular}

\footnotesize{
$^a$ The radial profiles are fitted with the analytical function $\rm \frac{y}{K/cm^{-3}} = \alpha + \beta \left( \frac{r}{pc} \right)$, where $y$ is the electron temperature or the electron density, $\alpha$ is the intercept of the radial profile and $\beta$ is the gradient of the radial profile. \\
}
\end{table*}\label{tab:fits}

\section{Comparison to models}\label{sec:model}

We combined stellar atmosphere models with the MAPPINGS V photoionization code to create a set of \hiireg\ region models.
The modeled \hiireg\ regions have a broad range of ionization parameter and pressure.

\subsection{Stellar ionizing source}

Ionizing sources of each \hiireg\ region are identified through the photometry of stars.
In LMC~N191, the dominant star is a $M_V=-5.27$~$mag$ O-star, of which the spectral type is classified as O5 V \citep{Selier-2012}. 
The ionizing source in SMC~N77A is suggested as an early-type B-star with an intrinsic color of B-V=-0.28 \citep{Selier-2012}.
\cite{Testor-2003} suggest that the ionizing source in SMC~N88A has a spectral type of O6-O8 V star with a $M_V$ of -4.85.
The stellar context in SMC~N456 is complex because it contains four bright O-type stars, with $M_V$ ranging from -4.5 to -4.75~$mag$ \citep{Testor-1987}.

We use the ionizing spectra from the {\sc TLUSTY} stellar atmosphere library \citep{Hubeny-1995,Hubeny-2017} for our photoionization models.
The {\sc TLUSTY} models are fully consistent, non-LTE (local thermodynamic equilibrium) line-blanketed atmosphere models allowing a broad range of effective temperature from 10\,K to $10^8$\,K.
We use the stellar atmosphere models with [Fe/H] abundance of -1.0 for the SMC nebulae and -0.7 for the LMC nebulae \citep{Cioni-2009}.

\subsection{Photoionization model grid}\label{sec:model_grid}

We used the MAPPINGS V photoionization code \citep{Binette-1985,Sutherland-1993,Dopita-2013} for modeling the ISM in compact \hiireg\ regions.
MAPPINGS V code utilizes the latest CHIANTI v.8.0 atomic database \citep{DelZanna-2015} for the lightest 30 elements.

MAPPINGS V is a photoionization code which is used to calculate the thermal and ionization structures of the ISM plasma with the equilibrium of cooling and heating, excitation and de-excitation. 
MAPPINGS V generates the nebular radial variation of the electron density, temperature, the ionic fraction and the emissions of 30 elements, under the assumption of the ISM density and pressure structure across the nebula.
MAPPINGS calculates the radiation from two components, the ionizing source and the ionizing recombination emission.
The fractional contribution of each component is determined based on the nebular geometry.

We construct the photoionization model grid with a set of constant ISM pressures ranging from $\log (P/k)$=5 to $\log (P/K)$=9 in a step of 0.05 dex, a set of ionization parameters from $\log(Q)$=6.0 to $\log(Q)$=9.5 in a step of 0.125 dex, and a varying stellar effective temperatures from $T_{\rm eff}$=30\,000\,K to $T_{\rm eff}$=50\,000\,K in a step of 2500~K.
Compared with the isochoric model which assumes a constant ISM density, constant ISM pressure is a more realistic assumption to the majority of \hiireg\ regions where the heating and cooling timescales are longer than the sound-crossing time \citep{Field-1965,Begelman-1990}.

We use the spherical geometry models for our comparison.
MAPPINGS is designed to generate the plane-parallel and spherical nebular models. 
The plane-parallel models are usually applied to the nebula whose radius is far larger than its thickness, while the spherical models are used for the cloud powered by a point source at its center.
Our observed four \hiireg\ regions have a round shape, with both LMC~N191 and SMC~N88A appearing spherical.
Therefore the spherical geometry is more appropriate than plane parallel geometry for our HII regions.

The ISM in the Magellanic Clouds have a non-solar element ratio which cannot be simply derived by scaling the solar metallicity.
We used the measured ISM metallicity in Magellanic Clouds from \cite{Russell-1992}.
In general, the mean metallicity of the LMC is 0.2~dex lower than our Galactic ISM, while the SMC has a metallicity 0.6~dex lower.
We included 15 elements in our LMC model and 16 elements in our SMC model. 
The details of element abundances are listed in Table~\ref{tab:magellanic_z}

\begin{deluxetable}{lcccc}
\tablewidth{8truecm}
\tablecaption {The abundance of the LMC and the SMC from \cite{Russell-1992} }
\tablehead {
\colhead {Element} & \colhead {LMC} & \colhead {SMC} \\
\colhead {} & \colhead {log(X/H)} & \colhead {log(X/H)} \\
}
\startdata  
\ H  & 0.0  &   0.0    \\
\ He &  -1.05  &  -1.09  \\
\ C  &  -3.96  &  -4.24  \\
\ N  &  -4.86 &  -5.37  \\
\ O  &  -3.65  & -3.97    \\
\ Ne  & -4.39 & -4.73  \\
\ Na  & -4.85 & -5.92  \\
\ Mg & -4.53  & -5.01   \\
\ Al  &   --  & -5.60  \\
\ Si  & -4.28  & -4.69    \\
\ S  & -5.29  & -5.41   \\
\ Cl  & -7.23 & -7.30 \\
\ Ar & -5.71 & -6.29   \\
\ Ca & -6.03 & -6.16    \\
\ Fe  & -4.77  & -5.11    \\
\ Ni  & -6.04  & -6.14    \\  
\enddata
\end{deluxetable}\label{tab:magellanic_z}

\subsection{Iterative Searching process}

We take a three-step fitting process to determine the best-fit model of the global emission-line spectrum of each \hiireg\ region.
The global spectrum is integrated across the entire field-of-view.
First, we use a coarse model grid to fit the temperature-sensitive line ratios, \nii$\lambda\lambda$6548,84/\nii$\lambda$5755, \oiii$\lambda$5007/\oiii$\lambda$4363, to roughly estimate the temperature of nebula and derive the theoretical full hydrogen spectrum based on the temperature.
The selected temperature-sensitive line ratios are free from dust extinction.
The \nii$\lambda\lambda$6548,84/\nii$\lambda$5755 ratio traces the temperature of the low-ionization zone of the nebula.
The \oiii$\lambda$5007/\oiii$\lambda$4363 ratio traces the temperature of the high-ionization zone of the nebula.
The coarse model grid has 5$<logP/K<$ 9 in a step of 0.5dex, 6$<logQ<$9.5 in 0.5dex steps and 30\,000\,K$<T_{\rm eff}<$50\,000\,K in a step of 25000\,K.

Then, we perform the extinction correction by comparing the full hydrogen spectrum between the selected model and the observation.
The extinction correction based on the Balmer decrement of \ha/\hb=2.86, needs to assume the nebula is optically thick with a constant temperature of $10^4$\,K (the Case-B nebula).
Instead, we fit the full hydrogen spectrum by using the Hydrogen emission-lines simultaneously, which is free from the arbitrary Case~B assumption.
The observed fluxes and the best-fit model fluxes of the hydrogen spectrum are listed in Table~\ref{apdx:N191h} to Table~\ref{apdx:N456h}.

The extinction correction is taken by converting the intensity of emission-lines, $I(\lambda)$, to the intrinsic intensity, $I_0(\lambda)$, through the equation below.
\begin{equation}
   I_0(\lambda) = I(\lambda)\times10^{C_n \times f(\lambda)},
\end{equation}
where $C_n$ is the logarithmic nebula reddening constant and $A(\lambda)/AV$ is the attenuation curve proposed by \citep{Fitzpatrick-1999}.
$f(\lambda)$ is the extinction function normalized by \hb\, $f(\lambda)$ = $A(\lambda)/AV)/ (Av/A(H_{\beta}) - 1$, to avoid the difference between the stellar extinction function and the nebular extinction function.
 
Finally, we search for the best-fit model of the de-reddened emission-line spectrum by using the fine model grid described in Section~\ref{sec:model_grid}.
The best-fit model is identified with the minimal $\chi^2$ value, where $\chi^2$ is defined as below,
\begin{equation}
\chi^{2} = \sqrt{\sum_{\lambda} \left(\frac{F_{\lambda}^{Obs}-F_{\lambda}^{Mod}}{F_{\lambda}^{Obs}}\right)^2},
\label{eq:chisq}
\end{equation}
where $F_{\lambda}^{Obs}$ and $F_{\lambda}^{Mod}$ are observed and modeled fluxes used at each step.


\begin{table*}
\caption{LMC~N191 Hydrogen Spectrum De-Reddened and Residuals}
\label{apdx:N191h}
\begin{center}
\begin{tabular}{r*{7}c}
\hline
\hline
 Wavelength                          & Obs. Flux  &  Error         & De--Red & De--Red    & $|$ Diff. $|$  \\
 (\AA)                       & (\hb = 100)  & 1$\sigma$ \%   &    Obs. Flux  &   Model     &   \%   \\
\hline
3734.375   & 2.427      & 0.91    & 2.740    & 2.397     & 12.52      \\
3750.158   & 3.046      & 1.39    & 3.433    & 3.048     & 11.21      \\
3770.637   & 3.775      & 1.43    & 4.246    & 3.963     & 6.66       \\
3835.391   & 6.563      & 1.56    & 7.330    & 7.292     & 0.52       \\
3970.079   & 14.89      & 0.79    & 16.40    & 15.86     & 3.29       \\
4101.742   & 24.42      & 0.65    & 26.53    & 25.83     & 2.64       \\
4340.471   & 44.03      & 0.19    & 46.68    & 46.74     & 0.13       \\
4861.333   & 100.00     & 0.94    & 100.00   & 100.00    & 0.00       \\
6562.819   & 331.05     & 0.23    & 285.02   & 287.73    & 0.95       \\
8437.956   & 0.477      & 0.70    & 0.370    & 0.317     & 14.32      \\
8467.254   & 0.567      & 0.44    & 0.439    & 0.376     & 14.35      \\
8502.483   & 0.639      & 1.22    & 0.494    & 0.450     & 8.91       \\
8665.019   & 1.188      & 1.83    & 0.913    & 0.839     & 8.11       \\
8750.472   & 1.518      & 1.62    & 1.163    & 1.068     & 8.17       \\
8862.782   & 1.925      & 2.99    & 1.468    & 1.389     & 5.38       \\
\hline
\end{tabular}
\end{center}
\end{table*}%

\begin{table*}
\caption{SMC~N77A Hydrogen Spectrum De-Reddened and Residuals}
\label{apdx:N77Ah}
\begin{center}
\begin{tabular}{r*{7}c}
\hline
\hline
 Wavelength                          & Obs. Flux  &  Error         & De--Red & De--Red    & $|$ Diff. $|$  \\
 (\AA)                       & (\hb = 100)  & 1$\sigma$ \%   &    Obs. Flux  &   Model     &   \%   \\
\hline
3734.375                                & 2.445 & 2.05           & 2.558   &  2.401   & 6.14    \\
3750.158                                & 2.950 & 1.32           & 3.084   &  3.059   & 0.81    \\
3770.637                                & 3.739 & 1.02           & 3.906   &  3.980   & 1.89    \\
3835.391                                & 6.730 & 0.66           & 7.012   &  7.336   & 4.62    \\
3970.079                                & 14.78 & 0.62           & 15.32   &  15.97   & 4.24    \\
4101.742                                & 25.04 & 0.34           & 25.82   &  25.99   & 0.66    \\
4340.471                                & 43.30 & 0.37           & 44.23   &  47.00   & 6.26    \\
4861.333                                & 100.00 & 0.59          & 100.00  &  100.00  & 0.00    \\
6562.819                                & 369.47 & 0.26          & 350.20  & 284.45   & 18.77   \\
8437.956                                & 0.334 & 2.91           & 0.307   &  0.307   & 0.15    \\
8467.254                                & 0.423 & 2.13           & 0.389   &  0.365   & 6.17    \\
8502.483                                & 0.418 & 3.72           & 0.385   &  0.437   & 13.51   \\
8665.019                                & 0.826 & 1.32           & 0.758   &  0.817   & 7.78    \\
8750.472                                & 1.135 & 1.98           & 1.041   &  1.041   & 0.03    \\
8862.782                                & 1.547 & 0.94           & 1.418   &  1.354   & 4.51    \\
\hline
\end{tabular}
\end{center}
\end{table*}%

\begin{table*}
\caption{SMC~N88A Hydrogen Spectrum De-Reddened and Residuals}
\label{apdx:N88Ah}
\begin{center}
\begin{tabular}{r*{7}c}
\hline
\hline
 Wavelength                          & Obs. Flux  &  Error         & De--Red & De--Red    & $|$ Diff. $|$  \\
 (\AA)                       & (\hb = 100)  & 1$\sigma$ \%   &    Obs. Flux  &   Model     &   \%   \\
\hline
3734.375 & 2.010   & 1.04  & 2.591   & 2.392   & 7.68      \\
3750.158 & 2.531   & 0.79  & 3.253   & 3.050   & 6.24      \\
3770.637 & 2.994   & 0.78  & 3.832   & 3.970   & 3.60      \\
3835.391 & 5.696   & 0.59  & 7.196   & 7.324   & 1.78      \\
3970.079 & 13.20   & 0.32  & 16.23   & 15.95   & 1.73      \\
4101.742 & 22.89   & 0.16  & 27.38   & 25.97   & 5.15      \\
4340.471 & 42.41   & 0.27  & 48.23   & 47.13   & 2.28      \\
4861.333 & 100.00  & 0.36  & 100.00  & 100.00  & 0.00      \\
6562.819 & 471.06  & 0.21  & 327.41  & 283.07  & 13.54     \\
8437.956 & 0.727   & 0.70  & 0.316   & 0.300   & 5.06      \\
8467.254 & 0.881   & 0.26  & 0.381   & 0.355   & 6.82      \\
8502.483 & 1.040   & 0.70  & 0.446   & 0.426   & 4.48      \\
8665.019 & 1.832   & 0.25  & 0.759   & 0.795   & 4.74      \\
8750.472 & 2.398   & 0.23  & 0.977   & 1.012   & 3.58      \\
8862.782 & 3.147   & 0.13  & 1.255   & 1.316   & 4.86      \\
\hline
\end{tabular}
\end{center}
\end{table*}%

\begin{table*}
\caption{SMC~N456 Hydrogen Spectrum De-Reddened and Residuals}
\label{apdx:N456h}
\begin{center}
\begin{tabular}{r*{7}c}
\hline
\hline
 Wavelength                          & Obs. Flux  &  Error         & De--Red & De--Red    & $|$ Diff. $|$  \\
 (\AA)                       & (\hb = 100)  & 1$\sigma$ \%   &    Obs. Flux  &   Model     &   \%   \\
\hline
 3734.375 & 2.334  & 0.54  & 2.556  & 2.401  & 6.06   \\
 3750.158 & 2.989  & 0.84  & 3.269  & 3.059  & 6.42   \\
 3770.637 & 3.812  & 0.62  & 4.162  & 3.981  & 4.35   \\
 3835.391 & 6.726  & 0.47  & 7.307  & 7.339  & 0.44   \\
 3970.079 & 14.54  & 0.24  & 15.64  & 15.98  & 2.17   \\
 4101.742 & 24.20  & 0.13  & 25.76  & 26.00  & 0.93   \\
 4340.471 & 43.36  & 0.44  & 45.32  & 47.05  & 3.82   \\
 4861.333 & 100.00 & 0.43  & 100.00 & 100.00 & 0.00   \\
 6562.819 & 341.59 & 0.43  & 304.33 & 283.79 & 6.75   \\
 8437.956 & 0.439  & 21.00 & 0.356  & 0.305  & 14.33  \\
 8467.254 & 0.542  & 1.11  & 0.438  & 0.362  & 17.35  \\
 8502.483 & 0.577  & 13.50 & 0.466  & 0.434  & 6.87   \\
 8665.019 & 0.976  & 4.73  & 0.783  & 0.811  & 3.58   \\
 8750.472 & 1.403  & 0.51  & 1.123  & 1.032  & 8.10   \\
 8862.782 & 1.343  & 2.36  & 1.070  & 1.344  & 25.61  \\
\hline
\end{tabular}
\end{center}
\end{table*}%

\begin{table}[!htbp]
\centering
\caption{Global Spherical Isobaric Derived Reddening Parameters.}
\label{tab:extparam}
\begin{tabular}{*{3}c}
\hline
\hline
                   & Rv            & $C_n$      \\
\hiireg\ Region     &               & (dex)     \\
\hline
{LMC~N191}       & 3.41  & 0.200          \\
{SMC~N77A}       & 2.74      &  0.061      \\
{SMC~N88A}       & 6.49      &  0.771       \\
{SMC~N456}       & 4.02      &  0.174     \\
\hline
\\
\end{tabular}
\end{table}

\begin{table*}
\centering
\caption{Global Spherical Isobaric Nebula Derived Properties}
\label{tab:1dfits}
\begin{tabular}{*{7}c}
\hline
\hline
                 & $\log (Q_H)$   & $T_{\rm src.}$ & $\log(P/k)$ & Fe Depl. & $<T_{\rm e}>$                & $<n_{\rm e}>$       \\
\hiireg\ Region     & (cm\,s$^{-1}$) & (K)             &     { }    & (dex)    & (K)                          & (cm$^{-3}$)         \\
\hline
{LMC~N191}       &  7.625         & 37\,500         & 6.80       & 0.5      & \multicolumn{1}{r}{ 9\,274}  & \multicolumn{1}{r}{339}    \\
{SMC~N77A}       &  8.125         & 38\,500         & 6.00       & 1.0      & \multicolumn{1}{r}{11\,860}  & \multicolumn{1}{r}{43}   \\
{SMC~N88A}   &  9.250         & 46\,500         & 7.40       & 1.5      & \multicolumn{1}{r}{14\,360}  & \multicolumn{1}{r}{1\,016} \\
{SMC~N456}       &  8.125         & 38\,500         & 6.45       & 1.5      & \multicolumn{1}{r}{12\,540}  & \multicolumn{1}{r}{114}    \\
\hline
 \end{tabular}
\end{table*}

\subsubsection{Best-fit model}

Table~\ref{tab:extparam} presents the $R_v$ and $C_n$ for each nebula.
The $R_v$ of SMC~N77A and LMC~N191 are consistent with the $R_v=$2.93 of the Small Magellanic Cloud and $R_v=$3.16 of the Large Magellanic Cloud \citep{Pei-1992} within an error of $\sim$6\%.
The $R_v$ of SMC~N88A and SMC~N456 are larger than the $R_v$ of the SMC.

Table~\ref{tab:1dfits} presents the derived parameters of the best-fit model.
These four nebulae have the central stellar temperature of 37\,500\,K$< T_{\rm eff} <$46\,500\,K.
The ionization parameter of the nebulae in SMC is 8.125$< log(Q_{\rm H}) <$9.250, which are larger than the typical ionization parameter 7$< log(Q_{\rm H} <$8 of local \hiireg\ regions \citep{Dopita-2000}.
The nebula in LMC has an ionization parameter of 7.625 falling in the range of ionization parameters of typical local \hiireg\ regions.
The ISM pressure is 6.0$< log(P/k) <$7.4 of these four nebulae.  

The electron temperature of nebula derived from the best-fit model is 9\,000\,K$<T_e<$14\,500\,K and consistent with the volume-weighted average temperature, $T_e$\oiii, measured from the IFU data.
The temperature of LMC~N191 is 9\,270\,K, matching with the measured temperature of 8\,864$\pm$813\,K. 
The temperatures of SMC~N77A and SMC~N456 are close to the typical assumption of $T_e=$10\,000\,K of an ideal ''Case B" nebula model.
The SMC~N77A has the temperature of $T_e=$11\,860\,K consistent with the measured temperature $T_e=$11\,405$\pm$630\,K.
The SMC~N456 has the derived temperature of $T_e=$12\,540\,K in agreement with the measured temperature $T_e=$11\,3215$\pm$423\,K.
The SMC~88A has the hottest temperature in the sample, with a derived temperature $T_e=$14\,170\,K consistent with the measured temperature of $T_e=$12\,932$\pm$1\,763\,K

The density of the nebula derived from the best-fit model is 10\cmcube$< n_e <$$10^3$\cmcube.
The SMC~N77A has the lowest density of $n_e=$43\cmcube.
The LMC~N191 and SMC~N456 have the density of $n_e=$339\cmcube\ and 114\cmcube\ respectively.
The SMC~N88A is the densest nebula with a density of $n_e=$1016\cmcube.

\subsection{Emission-line Ratios}

\begin{table*}
\centering
\caption{Key line-ratios of the nebulae and the best-fit models.}
\label{tab:keylineratio}
\begin{tabular}{*{16}c}
\toprule
\toprule
&{ } & \oiii/\oii & \oiii/\oiii & \oii/\oii & \sii/\sii \\
&{ } & (5007+4959)/(3726+3729) & 4363/(5007+4959) & 3726/3729 & 6716/6731 \\
\midrule
\midrule

&LMC-N191&   0.332  & 0.0035   &   0.90       &    1.15    \\
\midrule

&$\sigma$\%&  1.19   &  0.99  &    0.92    &     0.22      \\
\midrule

&Model &     0.291  &   0.0041  &   0.93       &    1.08   \\
\midrule

&$\Delta\%$&   12.3 &    17.1   &    3.3      &      6.1    \\
\midrule
\midrule

&SMC~N77A&    1.378  & 0.00754   &   0.724       &    1.344    \\
\midrule

&$\sigma$\%&  0.97   &  0.81  &    0.75    &     0.42      \\
\midrule

&Model &     1.179  &   0.00800  &   0.719       &    1.385   \\
\midrule

&$\Delta\%$&   14.45  &    6.13    &     0.71    &       3.07     \\
\midrule
\midrule

&SMC~N88A&    15.144  & 0.01415   &   1.236       &    0.875    \\
\midrule

&$\sigma$\%&  1.86   &  0.34  &    1.84    &     0.5      \\
\midrule

&Model &     13.499  &   0.0140  &   1.232       &    0.849   \\
\midrule

&$\Delta\%$&    10.86 &    0.71     &     0.32     &   2.97     \\
\midrule
\midrule

&SMC~N456&    1.899  & 0.00795   &   0.759       &    1.298    \\
\midrule

&$\sigma$\%&  0.93   &  0.45  &    0.86    &     0.19      \\
\midrule

&Model &     1.604   &   0.00987  &   0.767       &    1.301   \\
\midrule

&$\Delta\%$&    15.55 &    24.04     &     1.13     &   0.2     \\
\midrule

\bottomrule
\end{tabular}
\end{table*}

Table~\ref{tab:keylineratio} presents the comparison of the key integrated line-ratios between the observations and the best-fit model.
We compare the ionization-parameter-sensitive line-ratio \oiii$\lambda\lambda$ 4959,5007/\oii$\lambda\lambda$3726,27 (the well-known $O32$ ratio), the temperature-sensitive line-ratio \oiii\ $\lambda$4363/\oiii$\lambda\lambda$4959,5007, and the density-sensitive line-ratios \oii$\lambda$3726/\oii$\lambda$3729 and \sii$\lambda$6716 /\sii$\lambda$6731.
We use $\Delta\%$ to measure the derivation of the best-fit line-ratios from the observation. $\Delta\%$ is described as below,
\begin{equation}
\Delta\% = \frac{\lvert F_{\lambda}^{Obs}-F_{\lambda}^{Mod}\rvert }{F_{\lambda}^{Obs}}\times100,
\label{eq:chisq}
\end{equation}
where $F_{\lambda}^{Obs}$ and $F_{\lambda}^{Mod}$ are observed and modeled fluxes.

The density-sensitive \oii-ratio and \sii-ratio are well-fitted with the $\Delta\%$ smaller than 10\%.
The goodness of the temperature-sensitive line-ratio fitting varies from the $\Delta\%$ of 0.71\% in SMC~N88A to the $\Delta\%$ of 24.04\% in SMC~N456.
The $\Delta\%$ of the \oiii/\oii ratio ranges from 10-15\%.

\begin{figure*}
  \centering
  \includegraphics[width=7in]{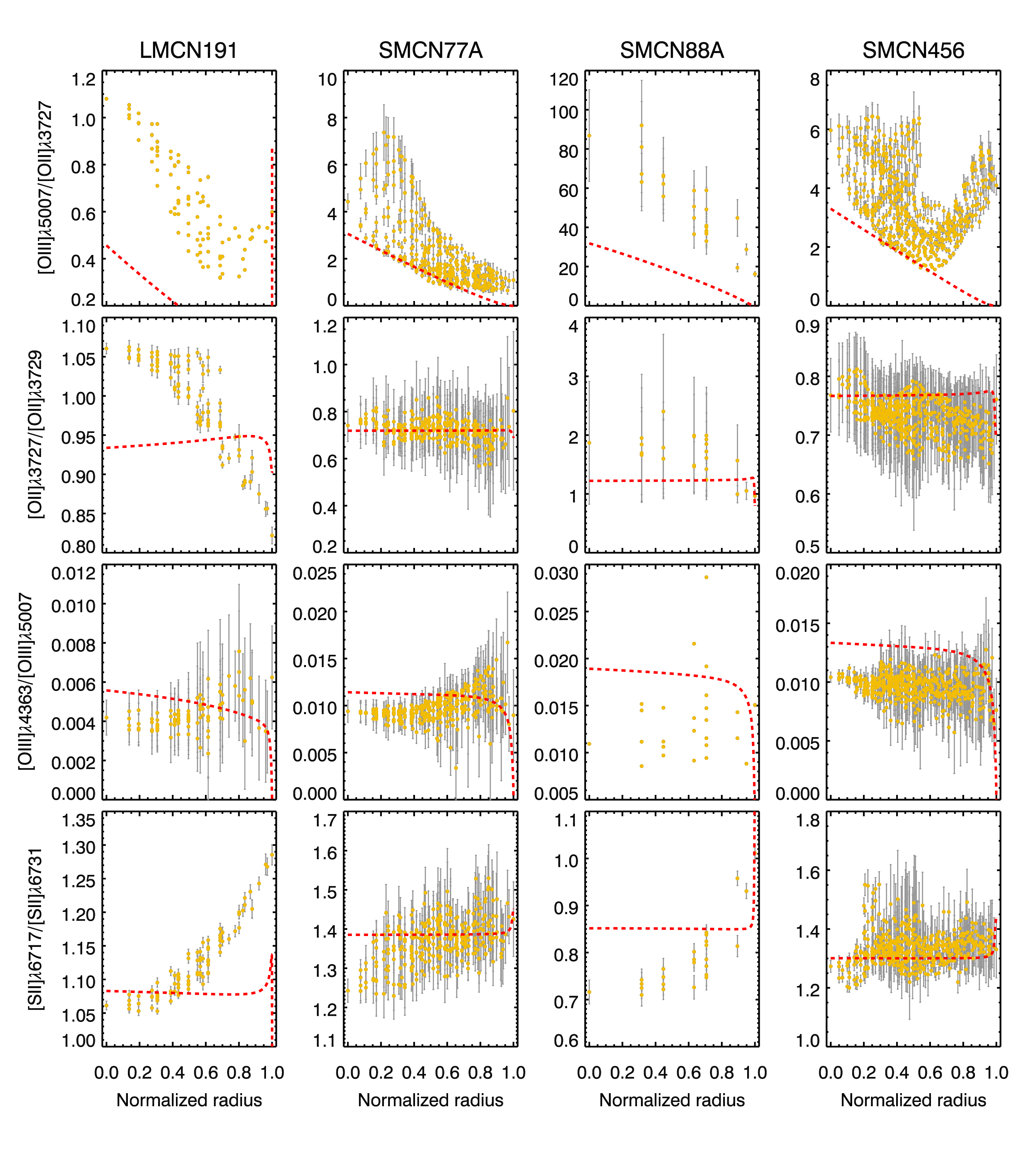}
  \caption{The comparison of radial distribution of the density-sensitive and the temperature-sensitive line ratios between our observational data and our constant pressure models. 
  Each row presents the radial distribution of each individual \hiireg\ region.
  The line ratios are the projection along the line-of-sight directions.
   The radius is normalized to the maximum radius of models for the constant pressure models and is normalized to the maximum radius of the nebula for the observed data.
{\bf Left:} The comparison of the radial distribution of the density-sensitive line ratio of \sii$\lambda$6716/\sii$\lambda$6731. 
{\bf Middle:} The comparison of the radial distribution of the density-sensitive line ratio of \oii$\lambda$3727/\oii$\lambda$3729. 
{\bf Left:} The comparison of the radial distribution of the temperature-sensitive line ratio of \oiii$\lambda$5007/\oiii$\lambda$4363. 
In each panel, the black points are the observed data and the lines present the radial distribution of \sii$\lambda$6716/\sii$\lambda$6731 given by constant pressure models with different log($P/k$) and ionization parameters.
}\label{fig:comp_data}
\end{figure*}

Figure~\ref{fig:comp_data} presents the comparison of the radial profile of the temperature-sensitive line-ratio, \oiii$\lambda$4363/\oiii\ $\lambda$5007, and the density-sensitive line-ratios, \oii$\lambda3727$\ /\oii$\lambda$3729 and \sii$\lambda6717$/\sii$\lambda$6731, between the best-fit model and the observation.
The distance to the center of nebula is normalized by the maximal radius of the nebula, $R_{max}$.
We project the 3D spherical nebula onto a 2D plane in order to compare with the observation.
The line-ratio at each distance is the integration along the line-of-sight direction.

The constant pressure model predicts that the \oii$\lambda3727$/\oii$\lambda$3729 ratio is an increasing function of the distance to the center of nebula.  
In contrast, the observation shows that the \oii-ratio decreases along the distance to the center of nebula for LMC~N191 and SMC~N456.
In SMC~N77A and SMC~N88A, the median value of the \oii-ratio displays a decreasing trend along the distance to the nebular center although the measurement error of 25-50\%\ on the \oii-ratio is too large to reveal an evident gradient of the radial profile. 

The observations show that the \sii$\lambda6717$ /\sii$\lambda$6731 ratio increases with the distance to the center of nebula for the four \hiireg\ regions. 
In contrast, the modeled \sii-ratio displays a flat distribution within the nebula.
The discrepancy of the \sii-ratio between the model and the observation reveals that the density structure of the nebula is more complex than the current models.

The temperature-sensitive \oiii$\lambda$4363/\oiii\ $\lambda$5007 ratio profiles are flat in the four \hiireg\ regions, which are consistent with the measurements of the flat temperature gradients in nebula.
In contrast, the constant pressure models predict a decreasing \oiii-ratio profile for the four \hiireg\ regions.

\subsection{The pressure of nebula}

\begin{figure*}
  \centering
  \includegraphics[width=7in]{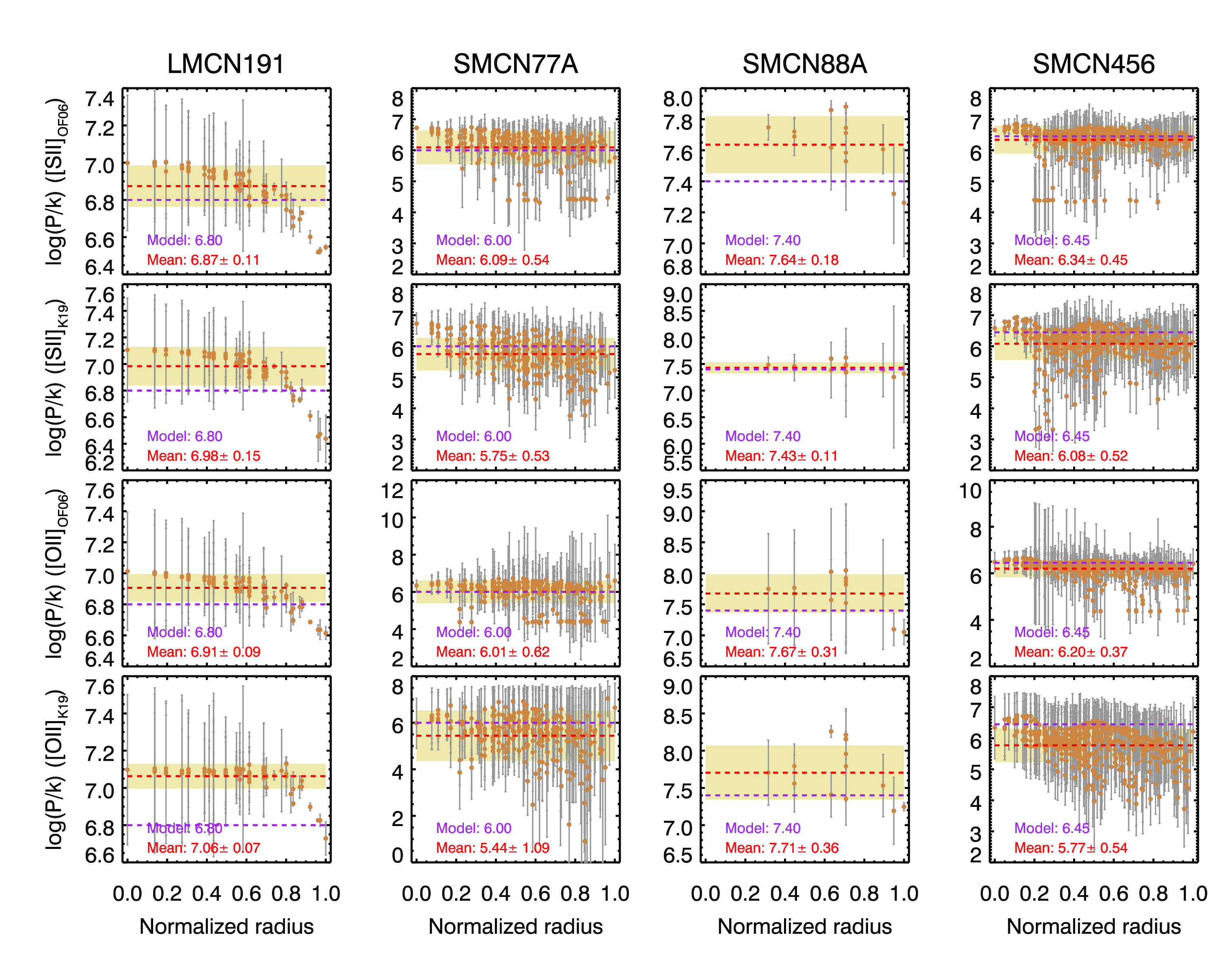}
  \caption{The ISM pressure profiles of the \hiireg\ regions. The pressure is derived through $P/k = 2 n_e (1+He/H)T_{e}$. The radius is normalized to the maximum radius of models for the constant pressure models and is normalized to the maximum radius of the nebula for the observed data.
  The red dashed line indicates the average of the derived pressure. 
  The purple dashed line indicates the modeled pressure value.
  The yellow region indicates the 1$\sigma$ of the measured pressure.}\label{fig:pres_prof}
\end{figure*}

Figure~\ref{fig:pres_prof} presents the radial profile of the pressure of the four \hiireg\ regions. 
For each nebula, the pressure, $log(P/k)$, is calculated by using four different indicators.
\begin{itemize}
\item $\log(P/k)^{OF06}_{\rm [SII]}$ is calculated by using $T_e$ measured by the \cite{Osterbrock-2006} method and $n_{e\rm [SII]}$ derived from \sii-ratio by using the \cite{Osterbrock-2006} method,
\item $\log(P/k)^{K19}_{\rm [SII]}$ is calculated by using $T_e$ measured by the \cite{Osterbrock-2006} method and $n_{e\rm [SII]}$ derived from \sii-ratio by using the \cite{Kewley-2019} method,
\item $\log(P/k)^{OF06}_{\rm [OII]}$ is calculated by using $T_e$ measured by the \cite{Osterbrock-2006} method and $n_{e\rm [OII]}$ derived from \oii-ratio by the \cite{Osterbrock-2006} method,
\item $\log(P/k)^{K19}_{\rm [OII]}$ is calculated by using $T_e$ measured by the \cite{Osterbrock-2006} method and $n_{e\rm [OII]}$ derived from \oii-ratio by the \cite{Kewley-2019} method,

\end{itemize}

The radial profiles of $log(P/k)$ shows an isobaric condition within the nebula.
In SMC~N77A, SMC~N88A and SMC~N456, the pressure has a flat distribution across the nebula.
In LMC~N191, the pressure has a flat gradient within 0.7$R_{max}$ and decreases from 0.7$R_{max}$ to 1$R_{max}$.
The average pressure of the nebula ranges from 5.4$\pm$1.1$< log(P/k) <$7.7$\pm$0.36, which is consistent with the range of pressures predicted by the best-fit models.

\section{Discussion}\label{sec:dis}

\subsection{Temperature Structures of \hiireg\ regions}


\begin{figure}
  \centering
  \includegraphics[width=\columnwidth]{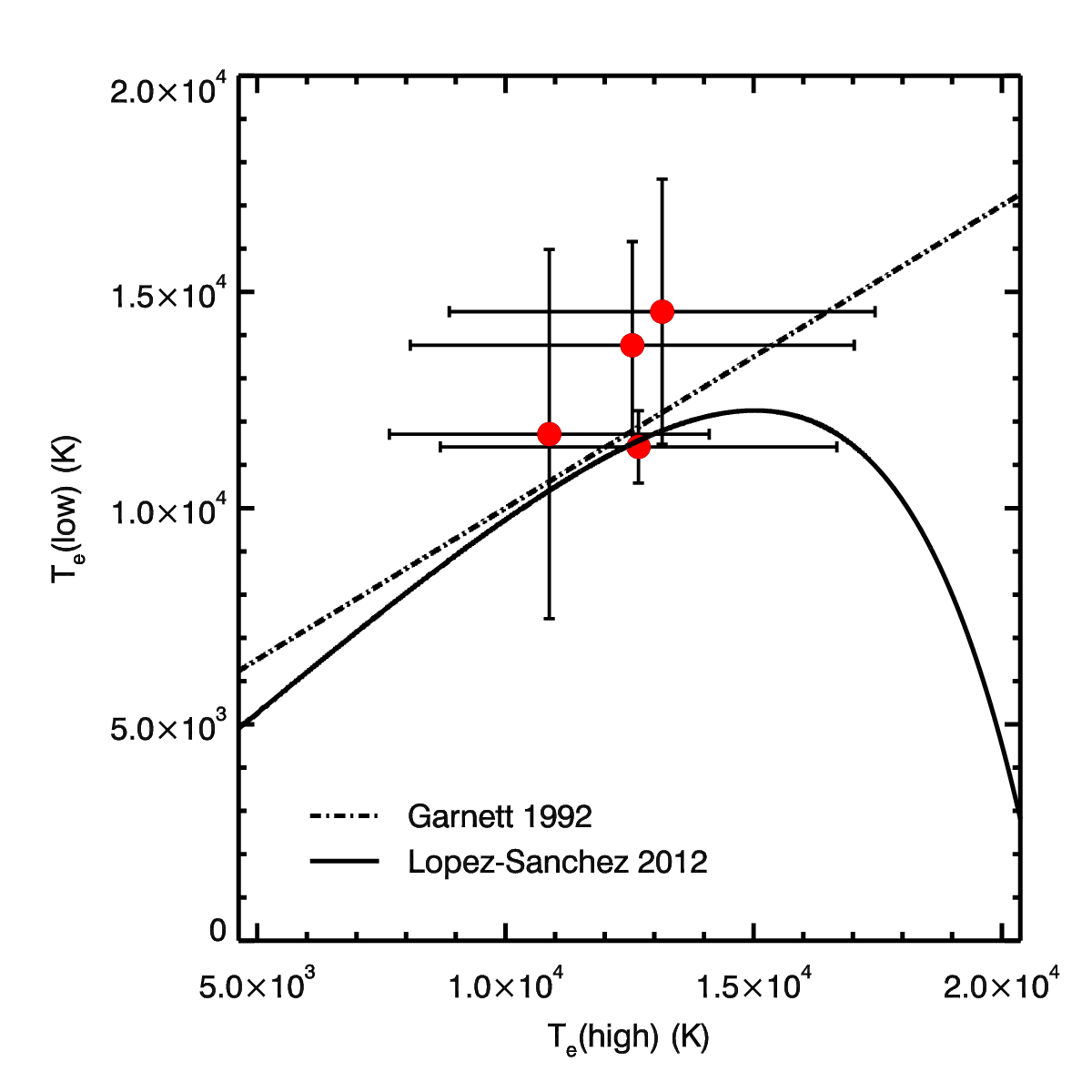}
  \caption{Comparison of the electron temperature of high-ionization species and the temperature of low-ionization species.
The temperature of high-ionization species is measured from the \oiii$\lambda$5007/\oiii$\lambda$4363 ratio.
The temperature of low-ionization species is measured from the \nii$\lambda\lambda$6548,84/\nii$\lambda$5755 ratio.
The solid line is the relationship given by \cite{Lopez-Sanchez-2012} and the dot-dashed line is the relationship induced by \cite{Garnett-1992}.}\label{fig:2modelfit}
\end{figure}


The electron temperature is a crucial diagnostic for determining the ISM metallicity.
Calibrations of metallicity are derived from simple photoionization models with constant temperature or density.
However, real \hiireg\ regions are composed of multi-ionization zones with fluctuations of temperatures and densities \citep{Garnett-1992}.
In our data, \hiireg\ regions have two distinct zones: a high-ionization zone and a low-ionization zone.
The high-ionization zone with high-ionization species, like $\rm Ne^{2+}$ and $\rm O^{2+}$, is located closer to the central star than the low-ionization zone with low-ionization species, like $\rm O^{+}$ and $\rm N^{+}$, within nebula.
This is because the stellar radiation field is absorbed in regions closer to the central star so that there are few ionizing photons to ionize atoms to high ionization stages in the outer regions of the nebula..

Analyses of integrated nebular spectra suggest that electron temperatures vary from a low-ionization zone to a high-ionization zone \citep{Peimbert-1967,Peimbert-2004,Peimbert-2017}.
The lines \oiii$\lambda4363$, \oiii$\lambda\lambda4959,5007$ are most often used for determining the temperature in \hiireg\ regions.
However, the temperature derived from the \oiii\ lines, $T_e\rm (OIII)$, only represents the temperature in the $\rm O^{2+}$ zones.
\cite{Hagele-2008} suggest that the missing $T_e$ in low-ionization zones can lead to an underestimate of the metallicity by 0.2~dex.
A two-ionization zone model is assumed to convert the \te\ in high-ionization zones to the \te\ in low ionization zones.
As shown in Figure~\ref{fig:2modelfit}, our \hiireg\ region sample suggests that the observed relationship of average temperatures in high-ionization zones and low-ionization zones matches with the analytical function given by previous research \citep{Garnett-1992,Lopez-Sanchez-2012}.

In contrast to the assumption of constant temperature in photoionization models, real \hiireg\ regions have radial variations of electron temperatures.
Detailed study of the $\theta^{1}$ Ori C area in the Orion nebula demonstrates that the electron temperature rises with distance from the ionizing star \citep{Rubin-2011}.  
Our data show a sign of the positive gradient of the temperature profiles within LMC~N191, SMC~N77A and SMC~N88A.
However, the uncertainties of the best-fit gradients are too large to exclude the possibility of a flat or negative gradient of the temperature profiles within these \hiireg\ regions.
SMC~N456 shows a sign of a negative gradient of the temperature profile but the uncertainty of the best-fit gradient is too large to exclude a positive or flat gradient.
The electron temperature gradient causes an underestimate of the derived metallicity calibration \citep{Stasinska-2005} and can cause the difference between metallicities calibrated from recombination lines and auroral lines \citep{Peimbert-2017}. 

The inhomogeneity of the electron temperature within \hiireg\ regions is attributed to several main mechanisms \citep{Peimbert-2017}.
One mechanism is the extreme density inhomogeneity leading to the low-temperature heavily shadowed area within \hiireg\ regions.
The second mechanism is the temperature inhomogeneity is the mechanical energy input by stellar winds and shocks \citep{Arthur-2016,O'Dell-2017}.
Another potential cause of temperature inhomogeneities is the spatial distribution of multiple ionizing sources within \hiireg\ regions.

\subsection{Density Structures of \hiireg\ regions}

Most \hiireg\ region models assume constant density across the nebula. 
However, the constant density assumption is not realistic given that real \hiireg\ regions have significant density variations.
The density variations are found to be 80-700~$\rm cm^{-3}$ in the Orion nebula \citep{Rubin-2011} and to be 40-4000~$\rm cm^{-3}$ within \hiireg\ regions in the inner region of the Milky Way \citep{Simpson-2004}.

The density radial gradients are complex in nearby \hiireg\ regions.
The measurement of electron densities across the $\theta^{1}$ Ori C area shows that the density variation is not a monotonic function of distance to the ionizing star \citep{Rubin-2011}.
In large nearby \hiireg\ regions, the density gradients are observed to be flat \citep{Garcia-Benito-2010,Ramos-Larios-2010} while the density gradients appear to be negative in compact \hiireg\ regions \citep{Binette-2002,McLeod-2016}.
Particularly, ultra-compact \hiireg\ regions have the steepest density gradients among all categories of nebulae \citep{Kurtz-2002,Johnson-2003,Phillips-2007}.

Our observed \hiireg\ regions also show a large range of density gradients.
The density gradient is as steep as around $-700~\rm cm^{-3}~pc^{-1}$ in SMC~N88A.
However, the density gradients can also be as flat as $-3~\rm cm^{-3}$ $\rm pc^{-1}$ in SMC~N456.

Diverse density gradients of \hiireg\ regions are attributed to complex nebular geometry.
In the Orion nebula, complex density gradients are likely the result of a turbulent \hiireg\ region \citep{Arthur-2016,Ha-2021} with bar-like structures \citep{van der Werf-2013, Rubin-2011}, where the contamination of scattered light from dense clumps change the electron density distribution \citep{Kewley-2019b}.

There are two major processes proposed that create complex density structures within \hiireg\ regions.
One is the ``collect and collapse'' scenario \citep{Elmegreen-1995}, where dense clumps are formed when \hiireg\ regions expand into turbulent ISM.
The other is the ``radiation driven implosion'' model \citep{Bertoldi-1989}, where overdensities within \hiireg\ regions are amplified by the heating of hot stars.
Both processes create an overdensity of the ISM \citep{Walch-2015, Schneider-2016}.
The stellar wind is another potential cause for the complex geometries of \hiireg\ regions \citep{Park-2010}.

\subsection{Pressure structures of \hiireg\ regions}

The ISM pressure is a key parameter to describe the ISM properties, which includes both nebular temperature and density structures.
The ISM pressure is determined by the mechanical energy produced by stellar feedback, the strength and the shape of the radiation field.
Models with a constant ISM pressure are more realistic than models with a constant density or temperature, when the sound-crossing timescale is shorter than the heating and cooling timescale.
Previous research shows that the condition of constant pressure occurs within the majority of \hiireg\ regions \citep{Begelman-1990,Gutierrez-2010}.

The value of ISM pressure crosses a broad interval in \hiireg\ region models \citep{Kewley-2019,Kewley-2019b}, ranging from a low pressure of log($P/k$)=4 to a high pressure environment with log($P/k$)=8-9 which is observed in high-redshift \hiireg\ regions \citep{Lehnert-2009}.
Our sampled \hiireg\ regions in the LMC and the SMC present ISM pressures around log($P/k$)=6-8, which is consistent with the range given by models.

\subsection{Implications for future photoionization models}

Self-consistent photoionization models, like {\sc cloudy} and {\sc mappings}, require as input an assumption of the shape of the nebula.
Constant temperature or density assumptions are always applied to simple photoionization models but this is not physically realistic.
Our observations reveal that the substructures, as filaments or dense knots, exist within real \hiireg\ regions, which are caused by stellar winds, ISM turbulence and the nebular expansion.
Our observations indicate that the temperature and density structures do not obey the isothermal or constant density assumptions.
Gradients of temperature and density structures exist in the LMC and SMC nebulae.
Allowing temperature and density to vary results in a more physically realistic model.

To model realistic \hiireg\ regions, some approximations of nebular geometries are applied.
Some studies have used plane-parallel approximations to mimic the behavior of non-spherical structures in nebula \citep{Levesque-2010}, and some use concentric shells to deal with the shell-like structures within \hiireg\ regions \citep{Pellegrini-2020}.
In these models, the contribution of the diffuse ionizing photons to the ionization field is assumed as a fixed fraction, which may bear no relation to a real nebula with complex geometries.

Fully self-consistent three-dimensional photoionization codes are needed.
Monte-Carlo radiative transfer techniques offer the promise of a substantial improvement over current simple structures to handle the complex geometry of nebulae.
Some Monte-Carlo photoionization codes are already available \citep{Ercolano-2008,Vandenbroucke-2018} and have shown promise in modeling the ionized gas around young stars \citep{Law-2011} and nebulae with filamentary structures \citep{Ercolano-2012}.
For a precise modeling of nebular emission-lines and internal structures, a Monte-Carlo code with comprehensive considerations of atomic data and ISM microphysics, like cooling and heating processes, is under development.
The new code incorporates the Monte-Carlo radiative transfer technique into the existing {\sc mappings} photoionization code and will be presented in a forthcoming paper (Jin et al. in prep.)

\vspace{25pt}

This research was supported by the Australian Research Council Centre of Excellence for All Sky Astrophysics in 3 Dimensions (ASTRO 3D), through project number CE170100013. L.J.K. gratefully acknowledges the support of an ARC Laureate Fellowship (FL150100113).



\end{CJK*}
\end{document}